\documentclass[useAMS,usenatbib]{mn2e}

\usepackage{times,graphicx,amssymb,natbib}

\newcommand{\mearth}{M_{\rm \oplus}}
\newcommand{\mjup}{M_{\rm Jup}}
\newcommand{\msol}{M_{\rm \odot}}

\title[Population Synthesis of Self-Gravitating Disc Fragments]{Towards a
  Population Synthesis Model of Objects formed by Self-Gravitating
  Disc Fragmentation and Tidal Downsizing}
\author[Duncan Forgan and Ken Rice]{Duncan Forgan $^{1}$\thanks{E-mail:
dhf@roe.ac.uk} and Ken Rice$^{1}$ \\
$^{1}$Scottish Universities Physics Alliance (SUPA), Institute for Astronomy, University of Edinburgh, Blackford Hill, Edinburgh, EH9 3HJ, Scotland, UK}
\begin{document}

\date{Accepted}

\pagerange{\pageref{firstpage}--\pageref{lastpage}} \pubyear{}

\maketitle

\label{firstpage}

\begin{abstract}

\noindent Recently, the gravitational instability (GI) model of giant
planet and brown dwarf formation has been revisited and recast into
what is often referred to as the ``tidal downsizing'' hypothesis.  The
fragmentation of self-gravitating protostellar discs into
gravitationally bound embryos - with masses of a few to tens of Jupiter
masses, at semi major axes above 30 - 40 au - is followed by a
combination of grain sedimentation inside the embryo, radial migration
towards the central star and tidal disruption of the embryo's upper
layers.  The properties of the resultant object depends sensitively on
the timescales upon which each process occurs.  Therefore, GI followed
by tidal downsizing can theoretically produce objects spanning a large
mass range, from terrestrial planets to giant planets and brown
dwarfs.  Whether such objects can be formed in practice, and what
proportions of the observed population they would represent, requires
a more involved statistical analysis.

We present a simple population synthesis model of star and planet
formation via GI and tidal downsizing.  We couple a semi-analytic
model of protostellar disc evolution to analytic calculations of
fragmentation, initial embryo mass, grain growth and sedimentation,
embryo migration and tidal disruption.  While there are key pieces of
physics yet to be incorporated, it represents a first
step towards a mature statistical model of GI and tidal downsizing as
a mode of star and planet formation.

We show results from four runs of the population synthesis model,
varying the opacity law and the strength of migration, as well as
investigating the effect of disc truncation during the fragmentation
process.  We find that a large fraction of disc fragments are
completely destroyed by tidal disruption (typically forty percent of
the initial population).  The tidal downsizing process tends to
prohibit low mass embryos reaching small semimajor axis.  The majority
of surviving objects are brown dwarfs without solid cores of any kind.
Around forty percent of surviving objects form solid cores of order
5-10 Earth masses, and of this group a few do migrate to distances
amenable to current exoplanet observations.  Over a million disc
fragments were simulated in this work, and only one resulted in the
formation of a terrestrial planet (i.e. with a core mass of a few
Earth masses and no gaseous envelope).  These early results suggest
that GI followed by tidal downsizing is not the principal mode of
planet formation, but remains an excellent means of forming gas giant
planets, brown dwarfs and low mass stars at large semimajor axes.
\end{abstract}

\begin{keywords}

planets and satellites:formation,  stars:formation,
accretion:accretion discs, methods: numerical, statistical

\end{keywords}

\section{Introduction}


\noindent Traditionally, there have been two distinct models or
paradigms of planet formation in protostellar discs.  In the core
accretion (CA) paradigm \citep{Pollack1996,Hubickyj2005}, planetary
mass bodies are assembled in a bottom-up process.  The protostellar
disc initially contains a population of interstellar grains of size
$\sim 1 \mathrm{\mu m}$, which coagulate into larger particles.  As
the grains grow past a few cm in size, they settle to the midplane,
aggregating into planetesimals a few km in size.  The planetesimals
can grow via collisions into solid protoplanetary cores, which can
then accrete a thin gaseous envelope.

The fate of this solid core then depends on the mass it can reach - if
the core exceeds a critical core mass of around 10 Earth masses
\citep{Mizuno1980}, the envelope accretion process is no longer
thermally regulated, resulting in runaway growth of the atmosphere,
producing gas and ice giants.  Below this critical core mass, the gas accretion
remains slow, and terrestrial planets can be formed.  

Core accretion is certainly the most accepted model of planet
formation, and it is the most successful at explaining the majority of
the observed exoplanet population.  Despite this, predictions of deserts in the exoplanet distribution at short periods and low mass have not been borne out by observations, and the existence of highly compact systems such as Kepler-11 \citep{Lissauer2011} are not easily explained.

Questions also remain regarding certain stages of the core growth process.  There are
barriers to grain growth at various size scales.  The most famous is
the ``metre barrier'', which is in reality a mixture of inhibiting
factors that take effect for grain sizes as small as millimetres
depending on the local disc properties.  As grains increase in size,
their relative velocities increase, and the probability of continued
growth decreases \citep{Blum2008}.  Also, gas drag works most
efficiently on metre sized bodies, causing rapid radial drift and loss
of solid material onto the central star
\citep{Whipple1973,Weidenschilling1977}.  More recently, light has
been shed on the ``bouncing barrier'', which occurs at slightly lower
relative velocities, and also inhibits grain growth
\citep{Guttler2010,Windmark2012}.

Conversely, in the gravitational instability (GI) paradigm
\citep{Cameron1978,Boss_science}, planetary-mass objects are assembled
in a top-down fashion through fragmentation of the gaseous
protostellar disc.  This occurs during a relatively brief phase of the
disc's life where it is sufficiently massive to be self-gravitating,
and can become gravitationally unstable.  Gravitationally unstable
discs must satisfy the Toomre criterion \citep{Toomre_1964}:

\begin{equation} Q = \frac{c_s \kappa_e}{\pi G \Sigma} <
  1.5-1.7, \label{eq:Q}\end{equation}

\noindent where $Q$ is the Toomre parameter, $c_s$ is the local sound
speed of the gas, $\kappa_e$ is the local epicyclic frequency (equal
to the angular velocity $\Omega$ in Keplerian discs) and $\Sigma$ is
the disc surface density.  This is a linear stability criterion, and
for axisymmetric perturbations $Q<1$ for instability.  Equation
(\ref{eq:Q}) refers to non-axisymmetric perturbations, where the
critical $Q$ value has been established by simulations (see
e.g. \citealt{Durisen_review} for a review).

The onset of gravitational instability produces trailing spiral
density waves which heat the disc, increasing $Q$.  The density perturbations produced by the
spiral waves eventually decay, allowing the disc to cool and become
unstable once more.  As new spiral structure is generated by the
instability, this cycle of growth and decay introduces turbulence into
the gas, usually referred to as gravito-turbulence \citep{Gammie}.  If the disc's
angular momentum transport is locally determined, this turbulence can
be represented by a ``pseudo-viscosity'' $\nu$
\citep{Balbus1999,Lodato_and_Rice_04,Forgan2011}, and as such is
amenable to the \citet{Shakura_Sunyaev_73} $\alpha$-parametrisation:

\begin{equation} \nu = \alpha c_s H, \end{equation} 

\noindent where $H$ is the disc scale height.  This description
greatly simplifies the disc evolution to a viscous form (although
typically non-linear and only soluble numerically) such that, if we
assume that the disc is axisymmetric, then the surface density
$\Sigma(r,t)$ evolves according to
\citep{Lynden-Bell1974,Pringle1981a}


\begin{equation}
\frac{\partial \Sigma}{\partial t} = \frac{3}{r}
\frac{\partial}{\partial r} \left[ r^{1/2} \frac{\partial}{\partial r}
  \left( \nu \Sigma r^{1/2} \right) \right] - \dot{\Sigma}_{\rm wind},
\label{eq:dSigmadt}
\end{equation}

\noindent where $\dot{\Sigma}_{\rm wind}$ represents mass loss from
the disc due to a disc wind.  The disc can respond to the heating
caused by the instability via radiative cooling, reducing the local
sound speed and decreasing $Q$.  As self-gravitating discs will only exist in the early stages of star formation, we should also expect a mass loading term in these equations which can have significant consequences for the disc's evolution (see e.g. \citealt{Clarke_09}).

As a result of the competition between heating and cooling,
gravitationally unstable discs can maintain what is referred to as a
marginally unstable state \citep{Paczynski1978}, where the shock
heating and radiative cooling become balanced, $Q$ maintains a
self-regulated value around the critical value, and the disc becomes
quasi-steady.  If the disc is marginally unstable and in local
thermodynamic equilibrium, then we can solve for $\alpha$ \citep{Gammie}:

\begin{equation} 
\alpha = \frac{4}{9 \gamma(\gamma-1)\beta_c}, 
\end{equation} 

\noindent where we have defined $\beta_c$ as a dimensionless
expression of the local cooling time $t_{\rm cool}$:

\begin{equation} 
\beta_c = t_{\rm cool} \Omega. 
\end{equation}

\noindent For a self-gravitating disc to fragment, this
self-regulating mechanism must be broken.  The density perturbations
induced by the gravitational instability have the behaviour
\citep{Cossins2008,Rice2011}:

\begin{equation} 
<\frac{  \Delta \Sigma_{\rm rms}}{\Sigma}> \propto \frac{1}{\sqrt{\beta_c}} \propto \sqrt{\alpha}. 
\end{equation}

\noindent Therefore, to produce a density perturbation of sufficiently
high amplitude that it can collapse into a bound object, the local
cooling time must be sufficiently short, or equivalently the local
gravitational stresses must be sufficiently high. 

It is reasonably clear that self-gravitating stresses saturate at some
value of $\alpha$, setting a lower limit for the cooling time, below
which a marginally unstable state can no longer be maintained.  Below
this critical cooling time $\beta_{\rm crit}$, the disc is no longer
able to achieve thermodynamic balance, and fragmentation is expected
to occur.  The critical cooling time is a function of the ratio of
specific heats ($\gamma$), but the saturating value of $\alpha$ is
fixed at around 0.06 \citep{Gammie,Rice_et_al_05}.

The precise saturating value of $\alpha$ is currently unclear due to
convergence issues \citep{Meru2011}.  Indeed, it may be the case that fragmentation  proceeds in a stochastic fashion, where the power spectrum of density fluctuations is such that a weak perturbation results in fragmentation, and the traditional fragmentation criteria do not delineate fragmenting/non-fragmenting configurations, but indicate where the probability of fragmentation is close to unity.  If this is the case, it may be that fragmentation can occur with low probability at values of $\beta_c$ as high as 20-50 \citep{Paardekooper2012,Hopkins2013}).  This debate is yet to be fully resolved, but there are indications that careful consideration of the effects of artificial viscosity
\citep{Lodato2011} and a more appropriate implementation of radiative cooling \citep{Rice2012} can help to address this problem, and that the saturating value of $\alpha$ remains of order 0.1. As yet, true stochastic fragmentation has not been observed in global disc simulations.  Due to our current ignorance of how its properties may be incorporated into this work, we will not include it in this analysis. 

Regardless, these arguments have to some extent excluded the effects
of mass infall, thermal history \citep{Clarke2007} and stellar
irradiation \citep{Kratter2011}.  Later work has generalised this
fragmentation criterion in terms of critical scales such as the Jeans
Mass \citep{Forgan2011a,Forgan2013} or the Hill
Radius \citep{Rogers2012}, which agree well with results derived from direct investigation from the spiral structure \citep{Boley2010b}.

Requiring the $Q$-condition for gravitational instability and a second
criterion for fragmentation restricts the disc radii at which
fragments may form.  Typically, the inner disc does not cool efficiently enough to maintain
low enough $Q$ values, and is therefore stable against self-gravity.
Also, gravitational stresses tend to decrease with increasing
proximity to the central star
\citep{Armitage_et_al_01,Rice_and_Armitage_09,Clarke_09,Rice2010}.

Fragmentation is therefore consigned to the outer disc, and typically
to radii above 30 or 40 au \citep{Rafikov_05, Matzner_Levin_05,
  Whit_Stam_06,Mejia_3,Stamatellos2008, intro_hybrid,
  Clarke_09,Kratter2009,Vorobyov2010,Forgan2011a}.  Estimates for the
fragment mass tend to place a lower limit of 3-5 $M_{\rm Jup}$
\citep{Kratter2009,Forgan2011a}.  These two results alone would appear
to be enough to discount the gravitational instability theory as a
means of forming lower mass planets at disc radii inside 30 au.
Equally, core accretion has difficulty explaining the existence of
massive planets with semi-major axes above 30 au, and so it may be the
case that planet formation proceeds via both core accretion and
gravitational instability \citep{Boley2009}.

However, the recent reformulation of the gravitational instability
theory into what is commonly referred to as ``tidal downsizing''
\citep{Boley2010b,Boley2011,
  Nayakshin2010b,Nayakshin2010,Nayakshin2010a} has revived the
possibility that the GI formation mode could produce low mass planets
at low semimajor axis.  The key to this revival is the study of the
subsequent evolution of the disc fragments into planetary embryos.
Three key physical processes shape the fate of the embryo:

\begin{enumerate}
\item \emph{Dust growth and sedimentation}.  As grains grow inside the
  embryo, the sedimentation velocity of the grains increases, drawing 
  them towards the pressure maximum at the embryo's centre. If this
  process is efficient, the gravitational field at the centre of the
  embryo will become dominated by solids.  The solids can become
  self-gravitating, potentially even gravitationally bound, forming a
  solid core.  As the temperature at the centre of the embryo
  increases, the grains can eventually be vapourised, ending the
  growth of the core.
\item \emph{Radial migration}.  Differential torques induced by the
  disc on the embryo can result in inward radial migration, in much
  the same way as protoplanets in the core accretion paradigm.
\item \emph{Tidal disruption}.  As an embryo migrates inwards, its
  physical radius will be comparable to the Hill Radius.  As this
  occurs, the upper layers will become unbound, and
  hence stripped from the embryo.
\end{enumerate}

\noindent By comparing the timescales on which these three processes
occur, \citet{Nayakshin2010} gives several possible outcomes for the
embryo.  In some cases, it is conceivable that a solid core
of less than ten Earth masses can be formed inside the embryo, which
has its upper layers completely disrupted, leaving a terrestrial
protoplanet.  Equally, inefficient core formation and strong tidal
disruption could result in the entire embryo being destroyed and
accreted by the star, giving rise to outburst behaviour
(e.g. \citealt{Vorobyov_Basu_05,Boley2010b,Dunham2012,Nayakshin2012a}).
Intermediate cases produce giant planets both with and without cores
depending on the sedimentation timescale.  If the core is not fully
gravitationally bound before the embryo is disrupted, then this could
potentially produce belts of planetesimals \citep{Nayakshin2012}, or generally deliver processed solid material that can affect the overall disc chemistry \citep{Boley2010b,Boley2011}.

This production of planets of varying types and masses, which
circumvents the need for planetesimal growth without precluding the
existence of asteroid belts, makes the tidal downsizing hypothesis an
attractive one for planet formation theorists.  However, it remains
unclear what distribution of object mass and semi-major axis should be
expected from self-gravitating disc fragmentation.  A common criticism
of gravitational instability theories as a whole is the lack of
population synthesis models, which have been implemented for the core
accretion model with a great deal of success
(e.g. \citealt{Ida_and_Lin_1,Alibert2005,Ida_Lin_mig,Mordasini2012}).

In this paper, we wish to remedy this by presenting a simple
population synthesis model for disc fragmentation followed by tidal
downsizing. This model is still missing some physical processes, and
simplifies others, but it gives the broad strokes of what stellar and
planetary systems we can expect to see produced by this theory.  As
such, we present it as the first stage in a continuing series of work
(much as was done with the population synthesis models of core
accretion).  We couple the equations of
\citet{Nayakshin2010,Nayakshin2010a,Nayakshin2010b} with semi-analytic
models of disc evolution incorporating X-Ray photoevaporation
\citep{Rice_and_Armitage_09,Owen2011}, and evolve a large ensemble of
planetary embryos formed in a variety of self-gravitating protostellar
discs.  While uncertainties remain about the underlying distributions
of disc mass and radius at early times, a uniform sampling of these
parameters allows us to explore the exoplanet parameter space in which
we can expect the GI formation mode to leave its signature.  This will
allow the statistical properties of the two principal planet formation
theories, core accretion and GI, to be compared much more directly
than has been possible until now.

The paper is structured as follows: in section \ref{sec:method} we
discuss the implementation of the equations that describe the
sedimentation, migration and disruption of embryos, as well as the
disc models used in this work.  In section \ref{sec:results} we
display the resulting statistical distributions as a function of a few
remaining free parameters. In section \ref{sec:discussion} we discuss
the implications of these models for the tidal downsizing hypothesis,
critically assess the current weaknesses of our approach and advocate
avenues for further work.  Finally, in section \ref{sec:conclusions}
we summarise the work.

\section{Method}\label{sec:method}

\subsection{A simple semi-analytic model of tidal downsizing}

\noindent In this section we summarise the equations governing embryo
formation, core formation and the subsequent migration/disruption
processes that can occur.  Throughout, we rely heavily on the
equations given by Nayakshin in his three papers outlining the tidal
downsizing hypothesis
\citep{Nayakshin2010,Nayakshin2010a,Nayakshin2010b}.  We also use
semi-analytic disc models as described in
\citet{Rice_and_Armitage_09}, and determine fragmentation and fragment
masses using the Jeans mass formalism as described in
\citet{Forgan2011a}.

\subsubsection{Conditions for Disc Fragmentation}

\noindent To firstly assess where fragmentation can occur, we
calculate the Toomre $Q$ criterion at all disc radii, as well as the
$\Gamma_J$ criterion outlined in \citet{Forgan2011a}, where

\begin{equation} \Gamma_J = \frac{M_J}{\dot{M}_J} \Omega. \end{equation}

\noindent and $M_J$ is the local Jeans mass inside a spiral
perturbation.  This condition allows us to determine at what rate the
Jeans mass is changing.  Regimes where the Jeans mass is decreasing
rapidly will be prone to fragmentation.  If we assume the disc is
marginally unstable and in a steady state, then

\begin{equation} \Gamma_J = \left(3/2\left(-\frac{1}{\beta_c} + \frac{9\alpha \gamma(\gamma-1)}{4}\right)\right)^{-1}. \label{eq:gammaj}\end{equation}  

 If the disc is in thermal equilibrium, then $\Gamma_J\rightarrow
 \infty$, and the disc will not fragment.  If $\beta_c$ is
 sufficiently low that thermal equilibrium would demand that $\alpha$ would exceed the value $\alpha_{sat}$ at which gravito-turbulent transport is expected to saturate, i.e. 
 $\alpha>\alpha_{\rm sat} \sim 0.1$, we set $\alpha=\alpha_{\rm sat}$
 and $\Gamma_J$ becomes negative.  As $\beta_c$ decreases, $\Gamma_J$
 remains negative, and its magnitude decreases.  Once $\Gamma_J$ is
 sufficiently small and negative, we can see that the local Jeans mass
 is decreasing sufficiently quickly that a local spiral density
 perturbation will soon exceed the Jeans mass, and begin the
 fragmentation process.

We therefore demand that (in disc regions where $Q\leq 1.5$) the local
Jeans mass must be decreasing at a rate such that

\begin{equation} -5 < \Gamma_J < 0 \end{equation}

\noindent for fragmentation to occur.  The exact critical values for
$\Gamma_J$ are currently unclear, but the above produce results
consistent with conventional cooling time criteria
\citep{Forgan2011a,Forgan2013}. The value of $\Gamma_J$ sets a timescale for
fragmentation rather than specifying whether or not it will occur.  if
$\Gamma_J < -5$, then in theory fragmentation remains possible, but
unlikely, as disc evolution will typically act to prevent this.

The advantage of using the Jeans mass criterion is that it immediately
identifies an initial fragment mass $M$.  From \citet{Forgan2011a}, this
is:

\begin{equation}M= M_J = \frac{4\sqrt{2}\pi^3 }{3G}\frac{Q^{1/2} c^2_s H}{\left(1 + \frac{\Delta \Sigma}{\Sigma}\right)}. \label{eq:mjeans_sigma}\end{equation}

\noindent While disc fragmentation is an inherently three dimensional process, the initial conditions provided by the disc are roughly two-dimensional.  If the surface density perturbation induced by a spiral density wave $\Delta \Sigma / \Sigma$ is of order unity, the perturbations will  become non-linear, and fragmentation can be expected to occur (this statement is true despite the current debate regarding stochastic fragmentation).  In other words, at the point of fragmentation, a 2D measure is sufficient to identify when fragmentation occurs, but the details of the fragment's subsequent evolution are 3D.  As the discs we use in the modelling done here are axisymmetric, surface density is in fact a 1D measure, and for these purposes is sufficient to control/govern the fragmentation process.  As with much semi-analytic modelling, using 1D measures as is done here is a necessary and appropriate sacrifice for the sake of computational expediency. 

Using the empirical relation for $\Delta \Sigma /\Sigma$ derived by \citet{Rice2011}, we can give the following form for $M$:

\begin{equation} M = \frac{4\sqrt{2}\pi^3 }{3G}\frac{Q^{1/2} c^2_s H}{\left(1 + 4.47\sqrt{\alpha}\right)}. \label{eq:mjeans_alpha} \end{equation}

\noindent Fragments are formed wherever both the $Q$ and Jeans mass
criteria specified above are satisfied.  We might expect fragments to
form at spacings of approximately 2 Hill radii, so that feeding zones
do not strongly overlap \citep{Lissauer1987}. To reflect our current
uncertainty in this matter, the precise value of the spacing is
uniformly distributed between 1.5 and 3 Hill radii.  This also allows
us to probe a larger variety of initial fragment properties
from what is a limited set of initial disc models.

This is sufficient to create the initial embryo population.  We must
now investigate the evolution of the embryos individually.

\subsubsection{Initial properties of the fragment/embryo}

\noindent By appealing to the analogy of the opacity limit for
fragmentation in star formation \citep{Rees1976,Low1976}, it can be
seen that fragments are initially akin to the ``first cores'' (FCs) of the
star formation process (e.g. \citealt{Masu_98}).  We will refer to
these as FCs to avoid confusion with the later solid cores that may
develop inside these embryos.

If these fragments are indeed FCs, it is appropriate to model them as
polytropic spheres.  The properties of the polytrope depend on the
initial temperature (taken usually to be $\sim 10 K$) and the critical
density $\rho_{\rm ad}$ at which the gas switches between isothermal
and adiabatic behaviour.  These quantities constrain the initial
radius of the FC \citep{Masu_98,Masunaga1999}.  If we follow the route
of \citet{Nayakshin2010} and define the opacity law

\begin{equation} 
\kappa(T) = \kappa_0 \left(\frac{T}{10K}\right)^{p_{\kappa}}, 
\end{equation}

\noindent and the following scaled quantities:

\begin{displaymath} m_1 = \frac{M}{0.01 M_{\rm
      \odot}}\end{displaymath}
\begin{displaymath} T_1 = \frac{T}{10 K} \end{displaymath} 
\begin{equation} \kappa_* = \frac{\kappa_0}{0.01},\end{equation}

\noindent we find the FC radius $R$ as (see \citealt{Nayakshin2010} for more):

\begin{equation} 
R = 17.5 \,\mathrm{au}\, m^{-1/3}_1 T_1^{(1+4p_{\kappa})/9}\kappa^{4/9}_* .
  \end{equation}

\noindent The temperature of the FC will be close to the virial
temperature:

\begin{equation} T_{\rm vir} = \frac{GM \mu}{3k_{B}R} = 146 \,\mathrm{K}\,
  m^{4/3}_1 T_1^{-(1+4p_{\kappa})/9}\kappa^{-4/9}_* .\label{eq:FC_T}\end{equation}

\noindent The luminosity of the FC is a non-trivial function of the
opacity power law index $p_{\kappa}$, and becomes somewhat cumbersome
to write down in general terms.  Hence, Nayakshin evaluates
expressions for $p_{\kappa} = 1,2$, and as such gives the equivalent
expressions for the cooling time for both opacity laws, assuming the binding energy of a
sphere:

\begin{equation} t_{\rm cool,0} = 380 \,\mathrm{yr}\, m^{2/3}_1
  T_1^{-4/3}\kappa^{1/9}_* \end{equation}

\begin{equation} t_{\rm cool,0} = 5700 \,\mathrm{yr}\, m^2_1
  T_1^{-19/9}\kappa^{-1/3}_* .\end{equation}

\noindent The ``0'' in the subscript indicates that these are initial
values for the cooling time.  Nayakshin shows that as the FC
contracts, the cooling time increases:

\begin{equation} t_{\rm cool}(t) = t_{\rm cool,0} +
  \left(1+p_{\kappa}\right) t .\end{equation}

\noindent As a result, the temperature evolves according to:

\begin{equation} T(t) = T(0)\left[1 +
    \left(1+p_{\kappa}\right)\frac{t}{t_{\rm
        cool,0}}\right]^{\frac{1}{1+p_{\kappa}}} .\end{equation}

\noindent The evolution of the temperature will be crucial to the
survival of grains inside the embryo, as will be discussed in the next
section.  Finally, we can express the contraction of the embryo as

\begin{equation} R(t) = \frac{R(0)}{\left[1+2\frac{t}{t_{\rm
          cool,0}}\right]^{1/2}} .\end{equation}

\noindent These analytic expressions should be contrasted with the more detailed calculations made by \citet{Helled2006} and subsequent works \citep{Helled2008,Helled2008a} which numerically integrate the equations of stellar evolution, with a spherically symmetric embryo as initial conditions.  Figure 1 of \citet{Nayakshin2010a} shows typical temperature evolution tracks using the above equations for FCs between 3 and 30 $\mjup$, and these compare well with the temperature evolution shown in Figure 2 of \citet{Helled2006}.  However, we should note that the contraction timescales shown here are much longer than those given by calculations where the opacity is allowed to vary due to grain evolution \citep{Helled2011}, and this has important implications for future work (see Discussion).

\subsubsection{Grain growth and sedimentation}

\noindent If there has not been significant grain growth or enrichment
before disc fragmentation, the protostellar disc grains (of no more
than a few microns in size) will be extremely slow to sediment to the
centre of the embryo.  Therefore, an initial phase of grain growth
must be completed before the embryo can form a core on a reasonable
timescale.  This grain growth timescale $t_{\rm gr}$ is initially
defined by the Brownian motion of the grains \citep{Dullemond2005},
but quickly becomes dominated by differential settling, where (at
relatively small grain sizes) the larger grains experience greater
sedimentation velocity.

Initially, the grains are small enough for their motion to be tightly
coupled to that of the gas.  The grain growth process is physically
very similar to that of the core accretion paradigm, with an added
gravitational acceleration towards the embryo's centre, which for a
constant density embryo is:

\begin{equation} a_{gr}(r) = -4/3 \pi G \rho r, \end{equation}

\noindent where $r=[0,R]$ is the grain's radial position.  The drag
force due to the embryo's gas can be expressed as a combination of
Stokes and Epstein regimes \citep{Weidenschilling1977,Boss1998,Nayakshin2010}:

\begin{equation} \frac{dv_s}{dt} = -\frac{\rho
    c_s}{\Sigma_s}\frac{\lambda}{\lambda + s}\left(v_s-v\right) - 4/3
  \pi G \rho r, \end{equation}

\noindent where $\lambda$ is the mean free path of the grain, $s$ is the
grain size, and $v_s$ and $v$ are the velocities of the grains and gas
respectively.  We have defined $\Sigma_s = \rho_{\rm bulk}s$, where
the bulk density of the grains $\rho_{\rm bulk}\approx 1
\,\mathrm{g\,cm^{-3}}$.  Grains will quickly reach a terminal velocity
$v_{\rm sed}$, which can be found by solving for $dv_s/dt=0$,

\begin{equation} v_{\rm sed} = \frac{-4\pi G \Sigma_s R}{3
    c_s}\frac{\lambda+s}{\lambda}. \end{equation}

\noindent Differential sedimentation allows the larger grains to sweep up the
smaller grains, and results in a mass growth rate:

\begin{equation} \frac{dm}{dt} = \pi s^2 f_g \rho v_{\rm
    sed}, \end{equation}

\noindent where $s$ is the grain size, $f_g$ is the mass fraction in
grains, and $\rho$ is the local gas density (i.e. in this initially
well-mixed phase, the volume density of grains $\rho_g = f_g \rho$).
Converting this into a growth rate for $s$ gives:

\begin{equation} \frac{ds}{dt} = \frac{f_g \rho}{4 \rho_s} v_{\rm sed}, \end{equation}

\noindent and so we can define a grain growth e-folding timescale:

\begin{equation} t_e = \frac{s}{\frac{ds}{dt}} = \frac{3 c_s}{\pi f_g
    \rho G R}. \end{equation}

\noindent Given an initial grain size ($s_0$) and final grain
size ($s_{\rm crit}$) before sedimentation can become significant, then the
growth timescale is

\begin{equation} t_{\rm gr} = \frac{3 c_s}{\pi f_g
    \rho G R} \ln \left(\frac{s_{\rm
      crit}}{s_0}\right). \label{eq:tgr0} \end{equation}

\noindent Once the grains have grown to $s_{\rm crit}$, the entire
body of grains (which we refer to as the ``grain sphere'') will begin
to sediment, quickly accelerating to critical velocity.  As the
sedimentation velocity is a function of $R$, the sedimentation time is
independent of radius, i.e. all shells of sedimenting dust will
reach the centre of the embryo at the same time (if the embryo is
uniform).  The sedimentation time is

\begin{equation} t_{\rm sed} = \frac{R}{v_{\rm sed}} =
  \frac{3c_s}{4\pi G \Sigma_s}\frac{\lambda}{\lambda+s}. \end{equation}

\noindent The sedimentation of grains can be limited by destructive
collisions \citep{Blum2008}.  At velocities of several metres
per second, collisions will result in fragmentation rather than
growth.  This limits the maximum value of $v_{\rm sed}$, and as such
the sedimentation timescale has a lower limit \citep{Nayakshin2010a}.  Generally speaking, these equations show similar dependencies on temperature and dust to gas ratio as those of Helled's calculations (see e.g. Appendix 1 of \citealt{Helled2011}), but we note that the physics we employ here is much simpler.

In particular, note that our equations describe the bulk motion of the grains, and do not incorporate the changing radial profile or temperature of the embryo.  The evolution of the clump's photosphere will have consequences that these equations cannot yet model.  Radiative heating of the embryo can have significant consequences for grain settling and destruction \citep{Helled2008a,Helled2011}.

\subsubsection{Turbulence as a suppressor of settling}

\noindent How should we select $s_{\rm crit}$ in equation
(\ref{eq:tgr0})? We have neglected turbulence in this analysis, which
will act to prevent gravitational settling \citep{Fromang2006}.  We
can approximate the effects of turbulent mixing using a 1D diffusion
equation in spherical geometry, with the diffusion coefficient
selected using the \citet{Shakura_Sunyaev_73} viscosity prescription:

\begin{equation} 
D = \alpha c_s R. 
\end{equation}

\noindent If the gas is initially a constant density sphere, the dust
sphere settles to an equilibrium solution

\begin{equation} 
\rho_{\rm s} = \rho_{\rm s,0} \exp \left[-\frac{R^2}{H^2_d}\right],
\end{equation}

\noindent where

\begin{equation} 
H_d = \sqrt{2Dt_{\rm sed}}.
\end{equation}

\noindent If we re-express this as an aspect ratio:

\begin{equation} 
\frac{H_d}{R} =\left[\frac{3\alpha}{2}\frac{\Sigma}{\Sigma_s}\frac{\lambda}{\lambda+s}\right]^{1/2}. \end{equation}

\noindent For gravitational settling to dominate over the turbulent
mixing the above aspect ratio should be much less than 1.  For a given
turbulence parameter $\alpha$, we can assume that $s_{\rm crit}$ is
found by setting $H_d/R = 1$ and solving for $s$.  Unlike the detailed calculations of \citet{Helled2008a} and \citet{Helled2011}, convection is not considered here.  They find that embryos are likely to be highly convective initially, and this has important consequences for slowing the settling process, and returning settled grains back to the upper envelope.  It is therefore crucial that future models incorporate convection (see Discussion for more). 

\subsubsection{The effect of the contracting embryo}

\noindent The grain growth timescale in equation (\ref{eq:tgr0})
tacitly assumes a static embryo.  However, the grain growth timescale
is sufficiently long compared to the cooling time that we should
expect the embryo to cool and contract significantly during this time.
This compression should accelerate grain growth.  The growth timescale
scales as

\begin{equation} t_{\rm gr} \propto \frac{c_s}{\rho R} \propto \left(\frac{T(t)}{T(0)}\right)^{-3/2}.
\end{equation}

\noindent Altering the size growth equation as follows:

\begin{equation} \frac{ds}{dt} =
  \left(\frac{ds}{dt}\right)_0\left[1+\left(1+p_{\kappa}\right)\frac{t}{t_{\rm
        cool,0}}\right]^{1.5/(1+p_{\kappa})}. \end{equation}

\noindent Integrating this gives the more accurate estimate of the
growth timescale:

\begin{equation} t_{\rm gr} = \frac{t_{\rm
      cool,0}}{1+p_{\kappa}}\left[\left(1+\frac{2.5+p_{\kappa}}{t_{\rm
          cool,0}}t_{\rm gr,0}\right)^\xi - 1\right], \end{equation}

\noindent where $\xi = (1+p_{\rm \kappa})/(2.5+p_{\rm \kappa})$. This
revised growth timescale can be over an order of magnitude shorter
than the static estimate, $t_{\rm gr,0}$.  Indeed, numerical
experiments indicate that the true timescale can be another factor of
two shorter still \citep{Nayakshin2010}, and as mentioned already, these timescales can be orders of magnitude shorter when opacity evolution due to grain growth is included \citep{Helled2011}.

\subsubsection{Core Formation}

The total mass in solids is 

\begin{equation} M_s = f_g M, \end{equation}

\noindent where $f_g=0.02$ for solar metallicity.  For an embryo of 1
$\mjup$, this gives a fiducial maximum core mass of around $60
\mearth$.  Approximately two thirds of the solids mass is ices
\citep{Nayakshin2010a}, so melting can reduce this maximum mass by the same
factor. 

As the sedimentation timescale is independent of radius, the grain
sphere, initially of radius $R_g(0)$ (which is equal to the radius of
the FC), will begin contracting according to

\begin{equation} 
R_g(t) = R \exp\left[-\frac{t}{t_{\rm sed}}\right] . 
\end{equation}

\noindent This allows us to identify the timescale $t_{\rm self}$ on
which the density of grains exceeds the local gas density, and the
grains become self-gravitating. This timescale is

\begin{equation} t_{\rm self} = t_{\rm sed} \frac{\ln
    f^{-1}_g}{3}. \end{equation}

\noindent The factor of 3 emerges from assuming a spherically
symmetric geometry.  Once the dust sphere is self-gravitating, the
sedimentation velocity is modified so that the gravitational
acceleration comes only from mass inside the dust sphere:

\begin{equation} v_{\rm sed} = -\frac{G
    M_s(R_s)}{R^2}\frac{\Sigma_s}{\rho c_s}\frac{\lambda
    +s}{\lambda} .\end{equation}

\noindent The mass interior to the dust sphere remains constant, and
therefore the differential equation for $R_s$ takes the form

\begin{equation} \frac{dR_s}{dt} = -\frac{A}{R_s^2}, \end{equation}

\noindent where $A$ is some constant, and has the solution:

\begin{equation} R(t) = R(t_{\rm self})\left[1-3\frac{t-t_{\rm
        self}}{t_{\rm sed}}\right]^{1/3}. \end{equation}

\noindent The contraction remains homologous, but is more rapid than
the non-self-gravitating stage.  The free fall time of the dust
sphere, i.e. the time required for all grains to collapse to infinite
densities, becomes approximately twice the sedimentation timescale
\citep{Nayakshin2010a}.

In this self-gravitating phase, the grain sphere can still be
dissipated.  If, for example, the outer envelope of gas is removed,
the resulting pressure gradient in the grain-gas mix can destroy the
grain sphere.  This is in part because the grain sphere is
self-gravitating, but not gravitationally self-bound.  If the grain
sphere can contract sufficiently to become self-bound, then even the
dissipation of the outer envelope will not unbind the sphere.
\citet{Nayakshin2010b} refers to this stage as the ``grain cluster''
stage, as an analogy to star clusters which survive interstellar gas
dispersal.

If the grain sphere's self-gravity dominates over the local gas pressure
gradient, then the grain sphere can enter the grain cluster phase.
This is in essence a Jeans length argument, and the usual calculations
apply.  The pressure gradient is of order

\begin{equation} \frac{dP}{dR} \sim \frac{\rho k_B T}{\mu m_H R}, \end{equation}

\noindent and should be equal to the gravitational force per unit
volume for hydrostatic equilibrium: 

\begin{equation} f_{\rm grav} \sim \frac{GM_{\rm
      enc}(R)}{R^2}\left(\rho +\rho_s\right). \end{equation}

\noindent Equating these (and assuming $\rho << \rho_g$) gives the Jeans length as

\begin{equation} \lambda_J = \left[\frac{3}{4\pi G}\frac{k_BT}{\mu m_H}\frac{\rho }{\rho^2_s}\right]^{1/2}. \end{equation}

\noindent Once the grain sphere radius contracts below this length,
then the final process of core formation can begin, which is presumed
to be rapid \citep{Boss1998,Nayakshin2010b}, although the process
requires further study to ascertain this (see Discussion).

\subsubsection{Ice Melting, Grain Vapourisation and Hydrogen Dissociation}

\noindent A fundamental barrier to solid core formation is the
increasing temperature of the embryo. As the embryo contracts, the
central temperature increases, and the chemistry and phase of the
grains will change.  At temperatures above $\sim 200K$, the ice
mantles of the grains will melt, preventing volatiles from becoming
part of the core. At temperatures above 1400 K, the grains can be
vapourised, preventing a solid core from forming at all.  Embryos of
relatively large initial mass will have higher initial temperatures,
and as a result will be more likely to vapourise their grains before a
core can form.

At temperatures above 2000 K, the molecular hydrogen inside the embryo
begins to dissociate.  The energy that must be stored in internal degrees of freedom to achieve this depletes the embryo's pressure support, allowing it to
begin a second collapse.  In star formation, this precipitates the
formation of the ``second core'' \citep{Masu_98}, the bound entity
which is the protostar itself.  Once this has occurred, the first core
equations no longer apply, and the embryo has attained a new
quasi-static configuration with a greatly reduced physical radius.

When objects reach this stage, we switch to the mass-radius model of
\citet{Burrows1997} (see e.g. \citealt{Burrows2011}).  We deem any
object which reaches this stage with a total mass greater than
$13\mjup$ to be a ``brown dwarf''.  Objects which reach this stage and
have a mass less than $13\mjup$ are split into several categories
depending on their final configuration.  Objects which have managed to
form a core before melting are recorded as ``icy core'' objects;
objects which form a core after melting are recorded as ``rocky core''
objects; and objects which do not form a solid core before the grains
vapourise are simply labelled ``no core'' objects.

From equation (\ref{eq:FC_T}), $T \propto M^{2/3}$, which gives a lower
mass limit for objects which are able to undergo $H_2$ dissociation
before being tidally disrupted.

\subsubsection{Disc Migration}

\noindent While the embryo is evolving internally, the disc is
evolving externally, and differential torques will result in planetary
migration.  The current wealth of literature on planetary migration
(see \citealt{Baruteau2012} and \citealt{Kley2012} for reviews) is
predicated on planets represented either as points or as uniform
density spheres.  The ``first core'' calculations shown above negate
both assumptions, as the embryo radius will typically exceed the Hill
radius, and as such more sophisticated hydrodynamic simulations are
required to show the migration of disc fragments in a disc which may
still be self-gravitating (e.g. \citealt{Baruteau2011,Michael2011,Cha2011,Boss2013}).  These simulations have shown that the migration of clumps in self-gravitating discs is non-trivial, and typically much more rapid than calculations based on non self-gravitating discs would predict.  Also, the migration direction becomes a function of the disc's viscosity, as has been found in other migration simulations (cf. \citealt{Bitsch2013}).

This being said, it is useful to use the standard migration timescale
calculations (as was done by \citealt{Nayakshin2010}) to allow
comparison with their work, as well as to keep the complexity of the
semi-analytic model under control.  Indeed, it is as yet unclear if a full
semi-analytic description of fragment migration has emerged from
numerical simulations, and will most likely require further study of
the parameter space before they can be formulated.

From \citet{Bate_cluster_03}, the migration regime depends on
the embryo mass.  Type I migration occurs if $M \leq M_t$, and Type II
if $M> M_t$, where the transition mass 

\begin{equation} M_t = 2 M_* \left(\frac{H}{R}\right)^3. \end{equation}

\noindent For Type I migration, the migration timescale is

\begin{equation} t_I(a) = \left(\frac{M}{M_*}\Omega\right)^{-1}
  \frac{H}{a}, \label{eq:typeI}\end{equation}

\noindent and for Type II

\begin{equation} t_{II}(a) = \frac{1}{\alpha \Omega}
  \left(\frac{H}{a}\right)^{-2} \label{eq:typeII} \end{equation}

\noindent (where we have assumed a self-gravitating, $Q=1$ disc).  As
expected, type I migration is typically several orders of magnitude
faster than type II.

Planet-disc interactions do more than modify the planet's semi-major
axis.  Other orbital elements can also be modified, principally the
eccentricity and inclination.  Co-orbital Lindblad resonances will
typically act to dampen eccentricities if the planet is fully embedded
\citep{Artymowicz1993}, but if the planet opens a gap dampening is
suppressed, and can even result in eccentricity growth depending on
the strength and saturation of corotation resonances
\citep{Ogilvie2003,Moorhead2008}.  In the case of massive planets
(i.e. above 1 $M_{\rm Jup}$), simulations indicate that eccentricity
is generally damped \citep{Bitsch2010,Kley2012}, but a multi-embryo environment may produce larger eccentricities through clump-clump scattering, or even eject them from the system entirely \citep{Boley2010b,Stamatellos2013}.  We assume that the
embryos adopt coplanar, circular orbits for simplicity.

\subsubsection{Tidal disruption}

\noindent If the embryo's radius exceeds its Hill Radius

\begin{equation} R_H = a\left(\frac{M}{3M_*}\right)^{1/3},\end{equation}

\noindent the upper layers can become unbound.  As the semi-major
axis decreases, the radial acceleration will typically increase, while
the embryo contraction rate decreases with time.  The inevitable
result is tidal disruption.  Appendix A of \citet{Nayakshin2010b}
indicates that while irradiation of the upper envelope by the star can
also achieve disruption, tidal disruption tends to occur first for
embryos around low mass stars (although the modification of the disc structure due to the radiation field, not modelled in this work, can affect the migration process, cf \citealt{Fouchet2012}).

The fate of the solid material inside the embryo depends on the
evolutionary stage.  As previously mentioned, if the grain sphere is
not self-bound, then it will become unbound as the envelope
disappears.  If the embryo is completely dissolved, the resulting
solids, which may have grown significantly since the embryo's birth,
are dispersed into the protoplanetary disc, as processed solids and rubble piles \cite{Boley2010b,Boley2011}, potentially even
planetesimals \citep{Nayakshin2012}.

If the grain sphere has become self-bound and initiated its final
collapse, then the stripping of the envelope can allow the emergence
of a terrestrial mass protoplanetary core.

\subsection{Evolution of the Protostellar Disc}

We assume that the disc is axisymmetric and that the surface density,
$\Sigma(r,t)$, evolves viscously according to equation
(\ref{eq:dSigmadt}). In a steady-state, this can be integrated to give
an expression for the mass accretion rate, $\dot{M}$, which, at radii
large compared to the radius of the star, is \citep{Pringle1981a}

\begin{equation}
\dot{M} = 3 \pi \nu \Sigma.
\label{eq:MassAccretion}
\end{equation}

One of the main uncertainties in disc evolution is what provides the
kinematic viscosity.  As discussed earlier, we assume that the
viscosity has the form of an $\alpha$ viscosity
\citep{Shakura_Sunyaev_73} so that $\nu = \alpha c_s H$, where $c_s$
is the disc sound speed, $H$ is the disc scaleheight ($H = c_s/\Omega$
with $\Omega$ the angular frequency of the disc), and $\alpha << 1$ is
a parameter that determines the efficiency of the angular momentum
transport.

We ran two sets of 100 disc models, selecting the central star mass
randomly between $M_{*} = 0.8$ M$_\odot$ and $M_{*} =
1.2$ M$_\odot$.  We use a ratioed grid that extends from $r=0.1$ au to
$r = 1000$ au.  

In one set of models, we assume the surface density of the disc model
is initially non-zero inside $r=50$ au only. When used in the population
synthesis model, we assume this disc initially extends to some value
of $r$ randomly selected between 50 and 100 au, and that the fragmentation
process truncates the disc to 50 au.

In the second set of models, the disc initially has non-zero surface
density inside its full 100 au extent, i.e. we assume that the
fragmentation process does not truncate the disc.

We assume an initial surface density profile of $\Sigma \propto
r^{-1}$.  Ideally, there should be a range of profiles measured: making the profile shallower may favour fragmentation by placing more mass at greater distance, which typically allows more rapid radiative cooling \citep{collapses}.  In this analysis, we decided to choose a single profile to keep the scope of the work under control, but future work should investigate this.

In each simulation, the initial disc mass is randomly chosen
to be between $0.125 M_{*}$ and $0.375 M_{*}$.  The
initial disc mass is therefore quite high and such discs are likely to
be self-gravitating.  Note that we do not add mass to the disc in any way - this is equivalent to the implicit assumption that the initial phase of cloud collapse and protostar formation is complete, and that the accretion of material onto the disc is much less than the mass loss due to winds.  This is an important assumption, as it curtails the accretion rates of the embryos, and future modelling work should address this.

Gravitationally unstable discs can either fragment to form bound
objects \citep{Boss1998,Boss2002} or they can settle into a
quasi-steady state in which the instability acts to transport angular
momentum \citep{Lin1987,Laughlin1994}.  

As made clear previously, it is now reasonably well established that
fragmentation is unlikely in protostellar discs within $20 - 30$ au.
If fragmentation does not take place, the gravitational instability is
likely, at early times at least, to transport angular momentum
outwards, allowing mass to accrete onto the central star.  In such a
state, the disc will achieve thermal equilibrium with dissipation due
to the gravitational instability balanced by radiative cooling.
Assuming $Q = 1.5$ therefore fixes the sound speed, and hence
temperature, profile in the disc.  It has been shown
\citep{Balbus1999,Lodato_and_Rice_04} that the gravitational
instability transports angular momentum in a manner analogous to
viscous transport.  The rate at which viscosity generates dissipation,
$D(R)$, in the disc

\begin{equation}
D(R) = \frac{9}{4} \nu \Sigma \Omega^2
\label{eq:diss}
\end{equation}

\noindent is balanced by radiative cooling, $\Lambda$, where \citep{Hubeny1990}

\begin{equation}
\Lambda = \frac{16}{3} \sigma \left(T_c^4 - T_o^4\right) \frac{\tau}{1
  + \tau^2}.
\label{eq:cool}
\end{equation} 

In Equation (\ref{eq:cool}), $\sigma$ is the Stefan-Boltzmann
constant, $T_c$ is the midplane temperature determined from $Q = 1.5$,
$T_o$ is a temperature due to external irradiation, and $\tau$ is the
optical depth.  We assume $\tau = \kappa \Sigma$, where $\kappa$ is
the opacity, and we use the analytic approximations from
\citet{Bell_and_Lin}. As described in detail in
\citet{Rice_and_Armitage_09} \citep[see also][]{Clarke_09,Zhu2009},
the above can be used to determine the effective value of $\alpha$.
Equation (\ref{eq:dSigmadt}) can therefore be integrated to determine
the time evolution of $\Sigma$.

If, however, the gravitational $\alpha$ is less than $0.005$, we
assume that another transport mechanism, such as the magnetorotational
instability (MRI) \citep{Balbus1991} will then dominate and we set
$\alpha = 0.005$. In this case, we no longer require that $Q = 1.5$
but set the midplane temperature, $T_c$, such that the cooling rate
given by Equation (\ref{eq:cool}) matches the dissipation rate given
by Equation (\ref{eq:diss}).  In addition, we also assume that
irradiation from the central star sets a minimum temperature in the
disc.

We also assume that radiation from the central star results in
photoevaporative mass-loss, producing a disc wind ($\dot{\Sigma}_{\rm
  wind}$).  There are a number of models for photoevaporative mass
loss, many of which assume that UV photons from the central star
ionise the disc surface
\citep[e.g.,][]{Hollenbach1994,Clarke2001,Alexander2006}.  Recent
models, however \citep{Owen2010}, suggest that x-rays are the dominant
driver and so here we implement the x-ray photoionization model
described in detail in \citet{Owen2011}.

Each disc model uses a different star mass, but assumes the same
disc-to-star mass ratio.  Despite this, the variation in x-ray
luminosity (from $5 \times 10^{28}$ erg s$^{-1}$ to $10^{31}$ erg
s$^{-1}$) produces a wide range of different disc lifetimes,
consistent with that observed \citep{Haisch2001}.

We therefore have a disc model that can self-consistently evolve the
surface density from the early stages when the gravitational
instability is likely to dominate through to the later stages when an
alternative transport mechanism, such as MRI, will dominate and also
includes the late-stage dispersal due to photoevaporative mass-loss.

\subsection{Implementation}

\noindent To conclude this section, we summarise the algorithm of the
semi-analytic model.

For each planetary system, we begin by randomly selecting a disc
model.  We either assume the disc truncates after fragmentation, or
does not.  If we are dealing with a truncated disc model, we initially
extend it to a maximum radius $r_{\rm max}$ beyond the model's outer
radius of 50 au. We sample $r_{\rm max}$ uniformly between 50 and 100
au, and we assume the disc surface density profile remains constant
in this extension, and calculate the subsequent disc properties
accordingly.  If we are dealing with a non-truncated disc model (which extends to 100 au), we do
not extend it in any way.  Again, $r_{\rm max}$ is somewhat arbitrary - to first order, increasing $r_{\rm max}$ will increase the number of fragments initially.  Whether this will increase the number of fragments that survive the tidal downsizing process is unclear.  Future work should investigate how changing this affects the resulting distribution of objects.  

The Jeans criterion is then used to find the minimum disc radius at
which fragmentation is expected to occur.  Embryos are then created,
with mass equal to the local Jeans mass, at spacings of a few Hill
radii (where the exact value is randomly sampled).  The other initial
embryo properties are calculated according to the first core
equations.  If we are assuming disc truncation, then we revert back to
the initial model, removing the material added to extend it to $r_{\rm
  max}$.  If we do not truncate the disc, then we make no
modifications at this stage.

The entire planetary system is then evolved on a timestep constrained
by the physical timescales in play.  The timestep can be no larger
than the minimum value of $t_{\rm gr}$, $t_{\rm sed}$ or $t_{\rm
  cool}$ for any embryos in the simulation.  The disc model is
interpolated from a look-up table to update its properties with time,
and these are used to calculate the migration timescales of
all embryos.  Note that each embryo is evolved independently of each
other - while several embryos may be evolved in the same disc in
parallel, they do not interact in any way, an important limitation (see Discussion).

Once an embryo's $t_{\rm gr}$ (which will be a function of $s_{\rm
  crit}$, itself a function of disc radius) has elapsed, sedimentation
is allowed to proceed, and the size of the grain sphere $R_s$ is
evolved (with the sedimentation velocity not allowed to exceed 10
metres per second due to fragmentation of the grains).  

At every timestep, the grain sphere is tested to check if it is
self-gravitating, or self-bound.  Grain spheres that are
self-gravitating use the accelerated sedimentation equation, and grain
spheres smaller than the Jeans length are considered to be self-bound,
and immediately form solid cores.  

Throughout this grain growth process, the embryo's central temperature
is monitored.  If the embryo exceeds temperatures of 180 K before core
formation, the ice mantles are melted, and two thirds of the solids mass
is deducted from the embryo.  If the temperature exceeds 1400 K, the
dust is vapourised and core formation is completely halted.  Embryos
with temperature above 2000 K enter the second core phase and are
assigned final radii according to the mass-radius relation of
\citet{Burrows1997}.

As the embryo migrates inward, it will begin the process of tidal
disruption.  Instead of immediately stripping the embryo, the embryo's
upper layers are removed more gradually over the course of one orbital
period, beginning at time $t_{\rm strip}$ with the radius decaying
exponentially to its Hill radius $R_H$ according to:

\begin{displaymath} R(t) = R(t_{\rm strip}) - \left(R(t_{\rm
    strip})-R_H\right) \end{displaymath}
\begin{equation} \left(1
  -\exp\left(\frac{\left(t-t_{\rm strip}\right)\Omega}{2\pi}\right)\right). \end{equation}

\noindent The mass $M$ of the embryo is then updated assuming the
embryo remains a polytrope of index 3/2.  If the new embryo radius is less than the
grain sphere radius, then the total mass in solids is also reduced by
an appropriate amount.

\section{Results}\label{sec:results}

\noindent To investigate the dependency on initial input
parameters, we run the population synthesis model with several
different configurations:

\begin{enumerate}
\item Setting $p_{\rm \kappa}=1$, and using disc models which
  initially extend to a radius randomly selected between 50 and 100
  au, and are then immediately truncated to 50 au after fragmentation
  has occurred.
\item As 1., except that the migration timescales are uniformly
  increased by a factor of 10, i.e. migration is ten times slower.
\item As 1., except $p_{\rm \kappa}=2$.
\item As 1., except the disc models extend to 100 au, and are not
  truncated after fragmentation.
\end{enumerate}

\noindent Every disc system is run for at least $1 Myr$, to ensure the full span of disc and embryo evolution is observed.

\subsection{Opacity Law Index $p_{\rm \kappa} = 1$ (Truncated Disc)}

\begin{figure*}
\begin{center}$\begin{array}{cc}
\includegraphics[scale = 0.4]{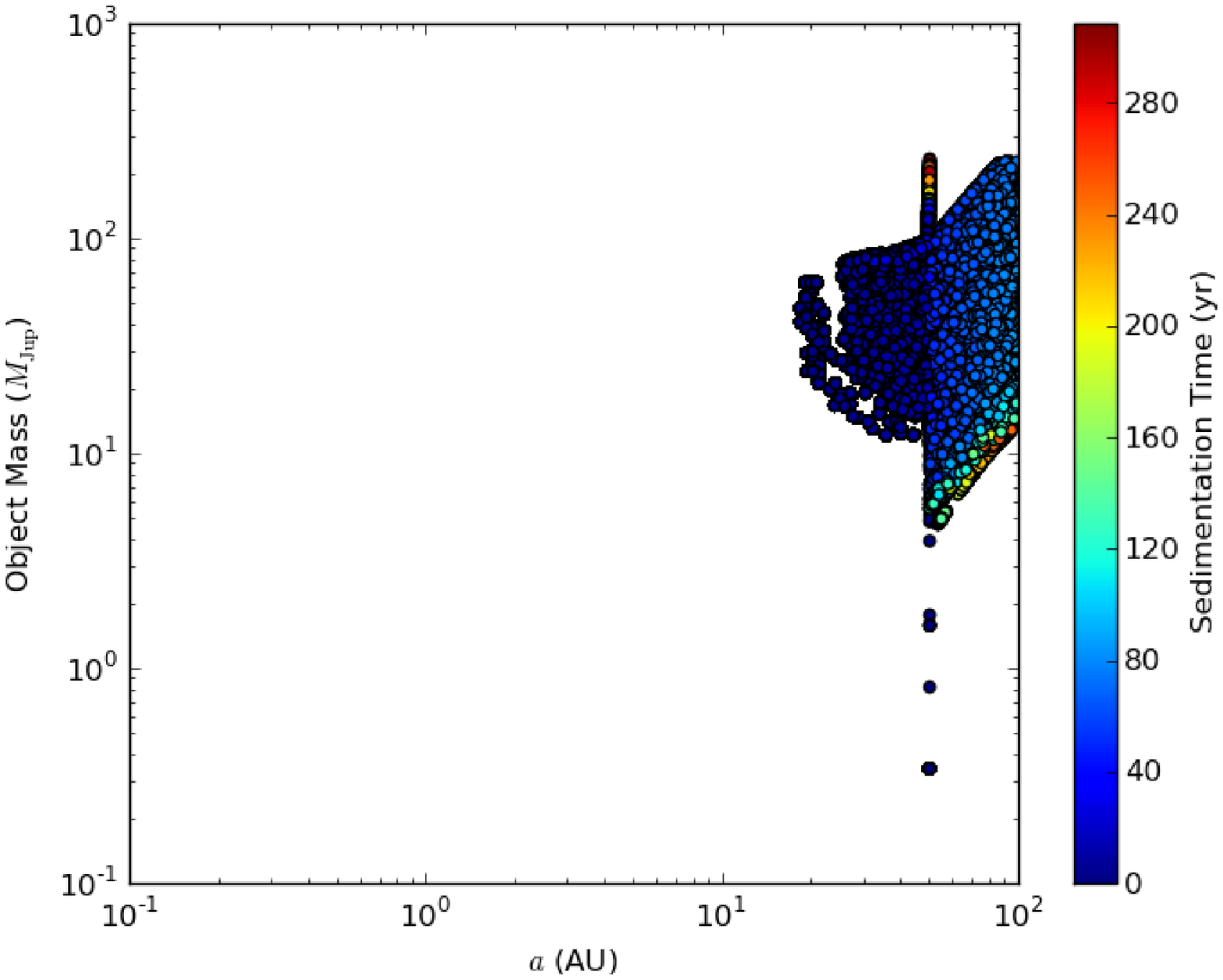} &
\includegraphics[scale=0.4]{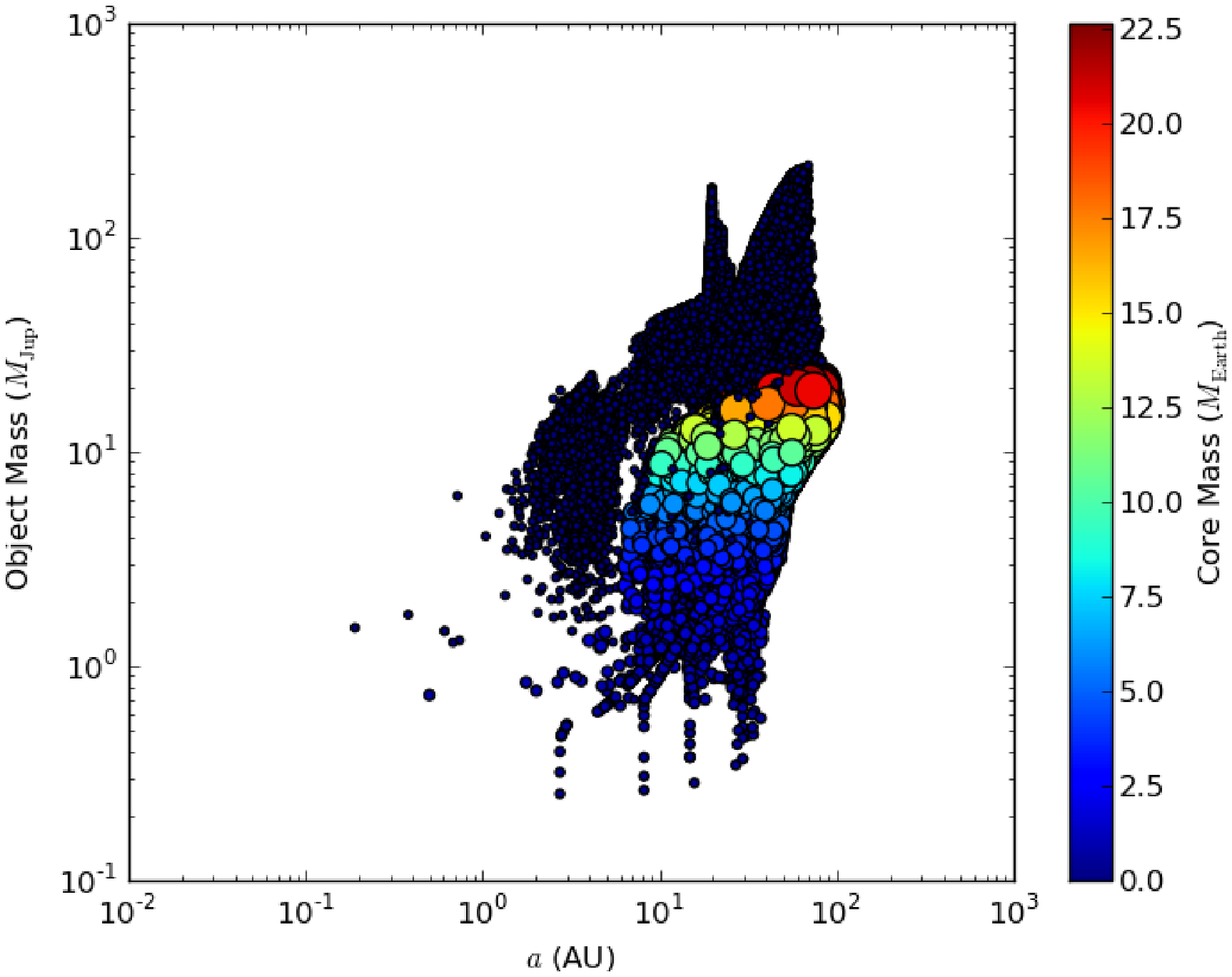} \\
\end{array}$
  \caption{Embryo mass versus semi-major axis for $p_{\kappa}=1$,
    where the disc truncates at 50 AU after fragmentation. The left hand panel
    shows the initial fragment mass-semimajor axis distribution, with
    the colour bar indicating the timescale on which grains will
    sediment once sufficiently grown.  256,424 disc fragments were
    produced in this run. The right hand panel shows the final
    distribution after $10^6$ years.  Of the fragments produced, only
    116,150 survived the tidal downsizing process, and 50,543 were
    able to form solid cores. The colours in this plot indicate the
    mass of the core in Earth masses. The disc model's inner boundary
    is fixed at 0.1 au, and as such objects cannot cross this
    boundary. \label{fig:m_vs_a_pkap1}}
\end{center}
\end{figure*}

\noindent The left panel of Figure \ref{fig:m_vs_a_pkap1} shows the
mass-semimajor axis distribution of the embryos at the point of
fragmentation.  As expected, the embryos initially occupy mass ranges
above a few $M_{\rm Jup}$, and fragmentation does not occur at small
$a$.  As some of the discs are quite massive and cool, the
fragmentation radius can be as low as 20 au.  The colour bar indicates
the grain sedimentation timescale for the embryos, which in the
majority of cases is relatively short ($<200 $ yr).  If the grains can
grow sufficiently quickly that sedimentation can begin before
vapourisation and disruption, core formation seems possible for many
of the embryos.

However, the rapid temperature evolution of the embryos, coupled with
tidal disruption beginning almost immediately after fragmentation,
tends to prevent core formation.  The right panel of Figure
\ref{fig:m_vs_a_pkap1} shows the mass-semimajor axis distribution of
the fragments formed in the population synthesis model after 1 Myr.
Note that most embryos do not possess a solid core, and are
sufficiently massive to be considered brown dwarfs.  The disc inner
boundary is located at 0.1 au, and as such all objects which reach
this location without being tidally disrupted are not evolved any
further.

\begin{figure*}
\begin{center}$\begin{array}{cc}
\includegraphics[scale = 0.4]{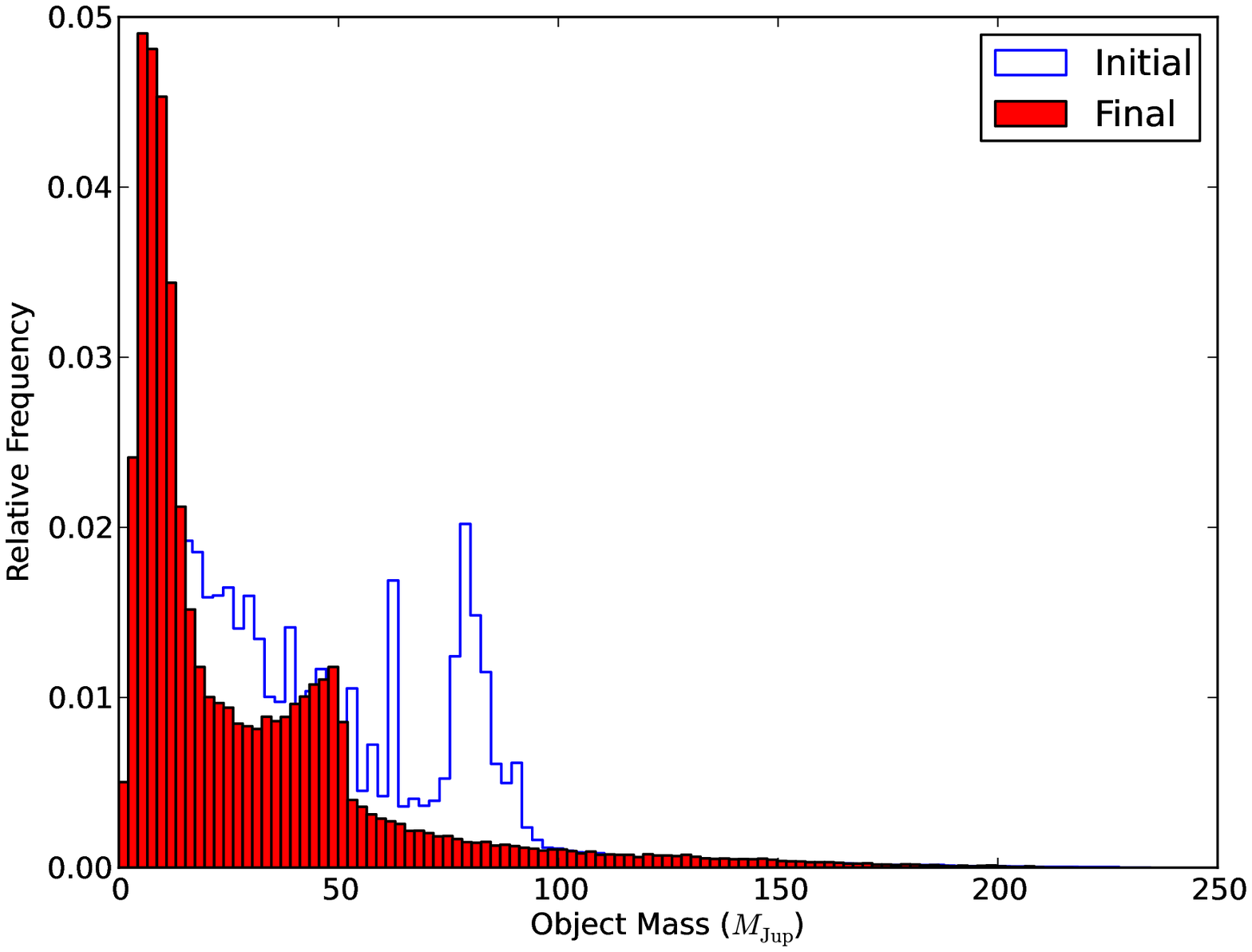} &
\includegraphics[scale = 0.4]{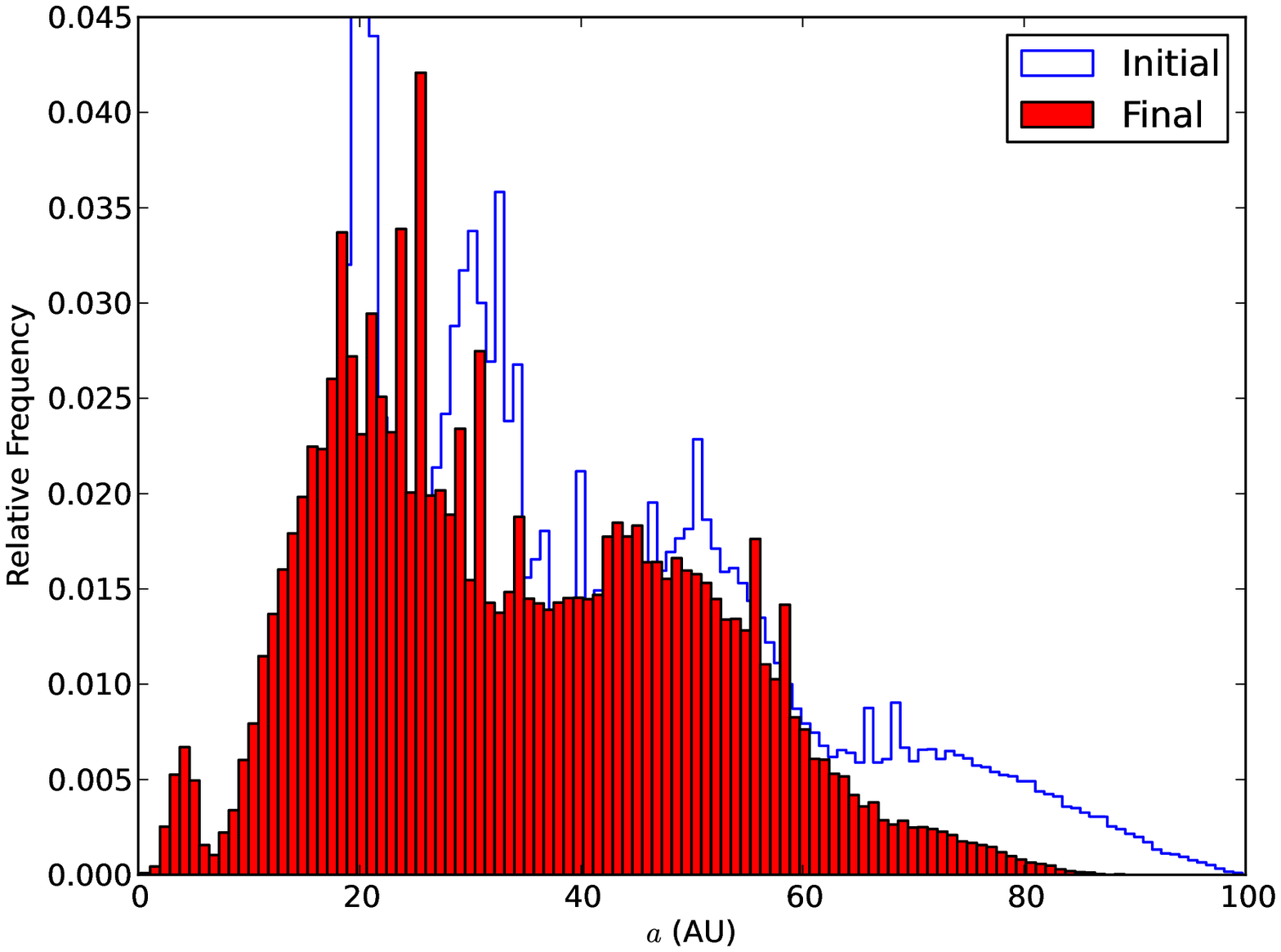} \\
\end{array}$
\caption{Comparing the initial (blue lines) and final (red bars) distributions of
  embryo mass (left) and embryo semimajor axis
  (right).\label{fig:compare_pkap1}}
\end{center}
\end{figure*}

There is a lack of objects at semimajor axes below $\sim 5$ au -
somewhat consistent with the ``brown dwarf desert''
\citep{Marcy2000,Halbwachs2000}) - which can be seen in the right hand
panel of Figure \ref{fig:compare_pkap1}.  Generally, objects above the
brown dwarf mass limit exist at semimajor axes $>20$ au, showing that
in situ formation of brown dwarfs via fragmentation remains viable in
the tidal downsizing paradigm, in agreement with recent numerical
simulations \citep{Stamatellos2009,Kaplan2012}.

If we consider both histograms in Figure \ref{fig:compare_pkap1}, we
can see that attrition of the embryo thanks to tidal disruption, and
the loss of embryos through migration strongly shapes the final
statistics of the objects.  Around 40\% of all embryos formed are lost
- as the mass of the embryo decreases during the disruption process,
the migration regime switches from Type II to Type I, significantly
reducing the migration timescale and forcing objects towards the
central star.  The mass distribution of objects (left panel of Figure
\ref{fig:compare_pkap1}) shows a significant number of objects formed
above $50 \mjup$, with masses in the hydrogen-burning M dwarf regime,
but the final mass distribution tends to fall off above this value as
disruption acts on these embryos before nuclear reactions begin.

While there are some embryos with final semimajor axes below $10$ au
(right panel of Figure \ref{fig:compare_pkap1}), the relative
frequency of objects is quite low.  Indeed, the final semimajor axis
distribution retains a strong signature of the minimum fragmentation
radius of about $20$ au.  Despite this fragmentation boundary being
much closer to the central star than the values of $40 - 70$ au
typically found in numerical simulations, the population of close-in
objects remains quite small compared to the large semimajor axis
population.  This appears to be due to disc truncation - as most of
the objects will form in regions that have experienced truncation,
they will not begin migration until the disc has spread on viscous
timescales.  This increases the migration timescale, and allows them
to evolve with limited tidal disruption and reach the ``second
collapse'' due to $H_2$ dissociation before the disc can act to move
the objects inward, and increase the rate of tidal disruption.


\begin{figure*}
\begin{center}$\begin{array}{cc}
\includegraphics[scale = 0.4]{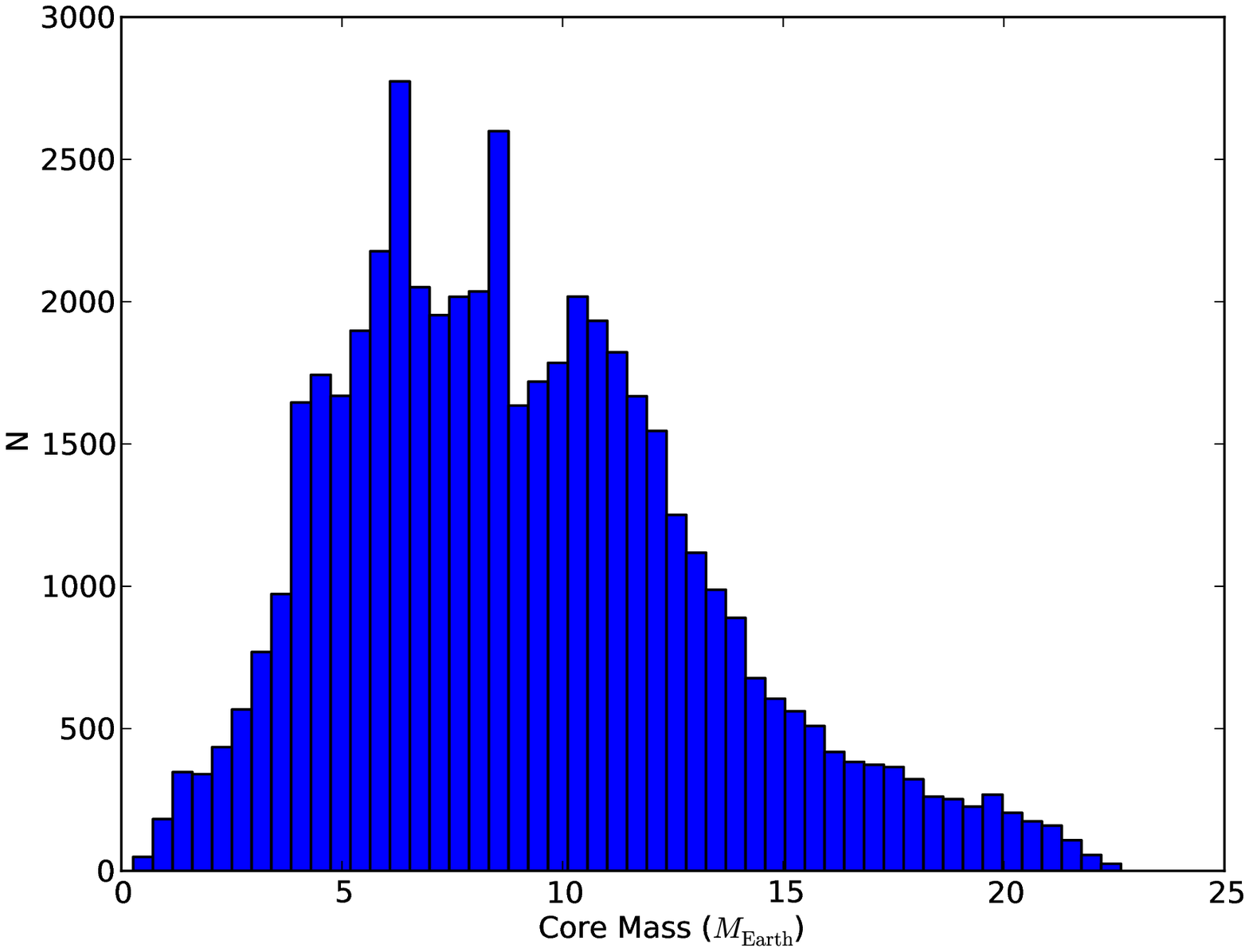} &
\includegraphics[scale = 0.4]{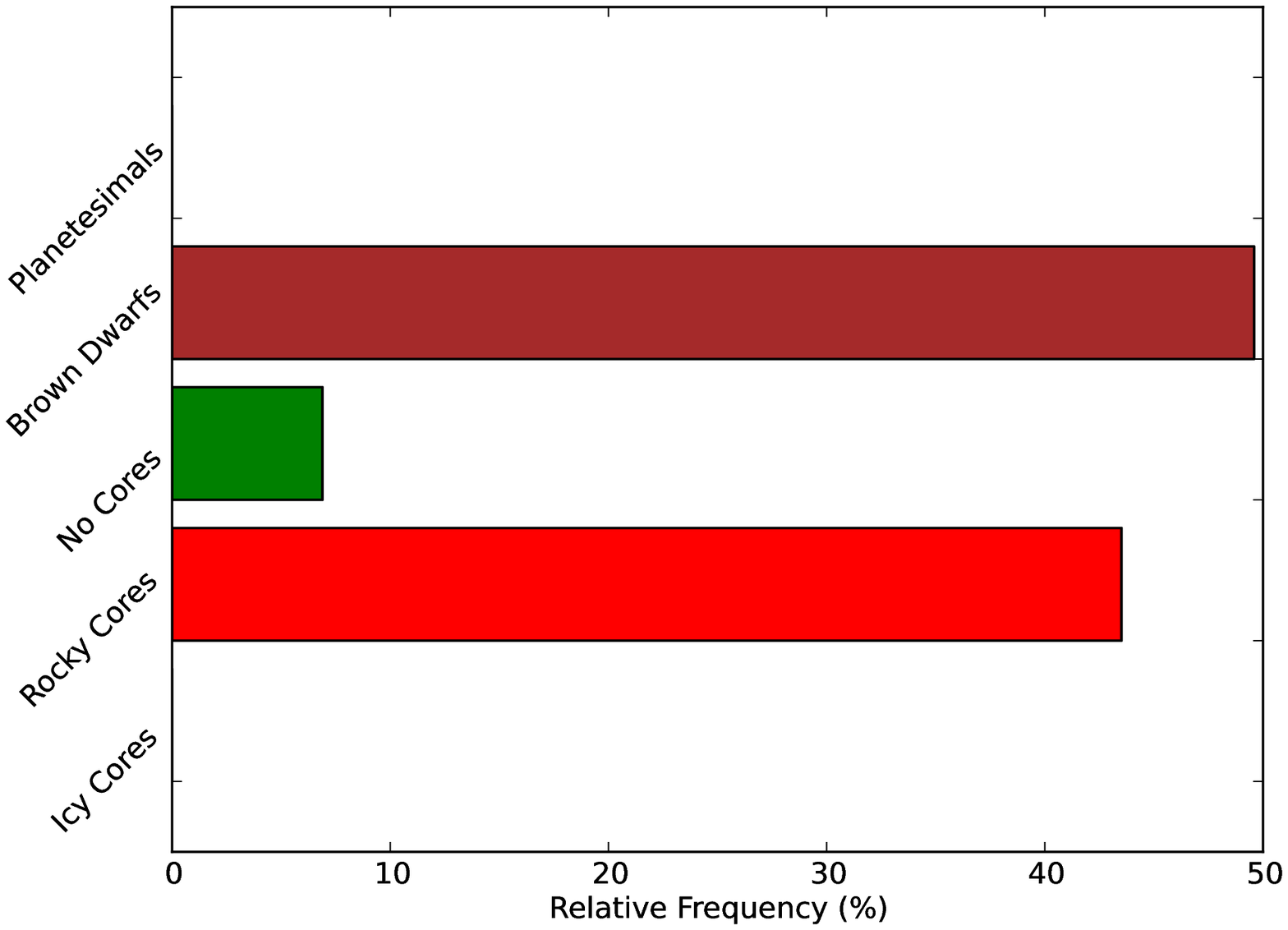} \\
\end{array}$
\caption{Left: The distribution of core masses after $10^6$
  years of evolution.  Right: The relative frequency of each object
  type formed in the population synthesis model.
  (right).\label{fig:mcore_types_pkap1}}
\end{center}
\end{figure*}

The left panel of Figure \ref{fig:mcore_types_pkap1} shows the
distribution of solid core masses formed in this run.  The
distribution is peaked at around $6\mearth$, with a tail extending as far as
$22 \mearth$.  The extrema of the distribution compares well to (for
example) the range of estimates for Jupiter's core mass (0-18
$\mearth$, \citealt{Guillot1999,Nettelmann2011}), but we must also
remember that objects formed in the model with these core masses are
likely to have total masses of $10\mjup$ or above.

The right panel of Figure \ref{fig:mcore_types_pkap1} illustrates that
brown dwarfs are the most populous of all the surviving objects in
this run (around 50\% of the total), closely followed by objects with
a solid core (at around 42\%).  No icy cores are formed, as the
embryos quickly exceed the melting temperature before core formation.
If and when the grain growth process is successful, the sedimentation
timescale is sufficiently short that cores are formed before full
disruption occurs.  We should therefore conclude that (for these parameters) these disruption events do not add significant processed materials or sub Earth mass solid bodies to the surrounding environment.

\subsection{Opacity Law Index $p_{\rm \kappa} = 1$ (Truncated Disc, Inefficient Migration)}

\begin{figure*}
\begin{center}$\begin{array}{cc}
\includegraphics[scale = 0.4]{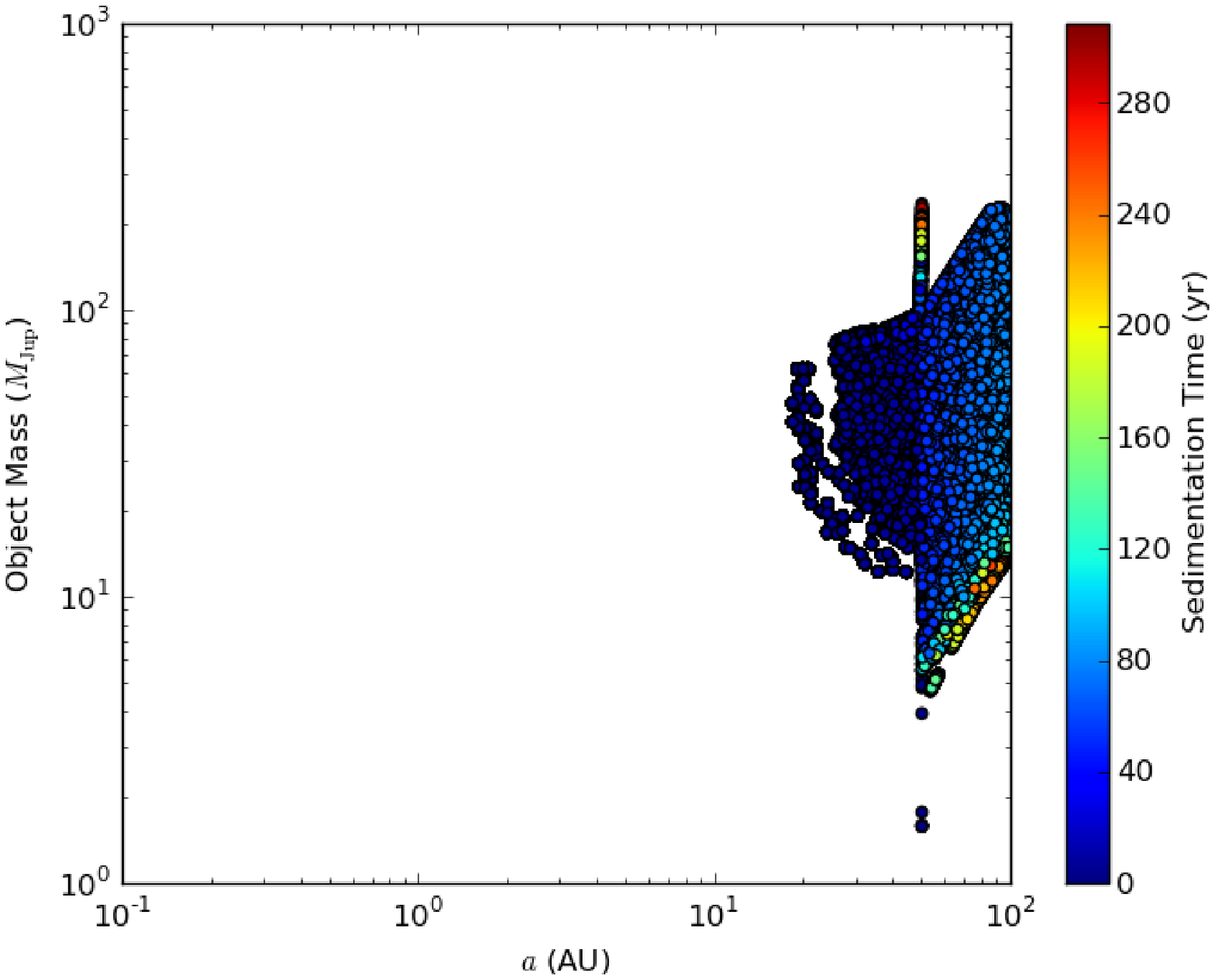} &
\includegraphics[scale=0.4]{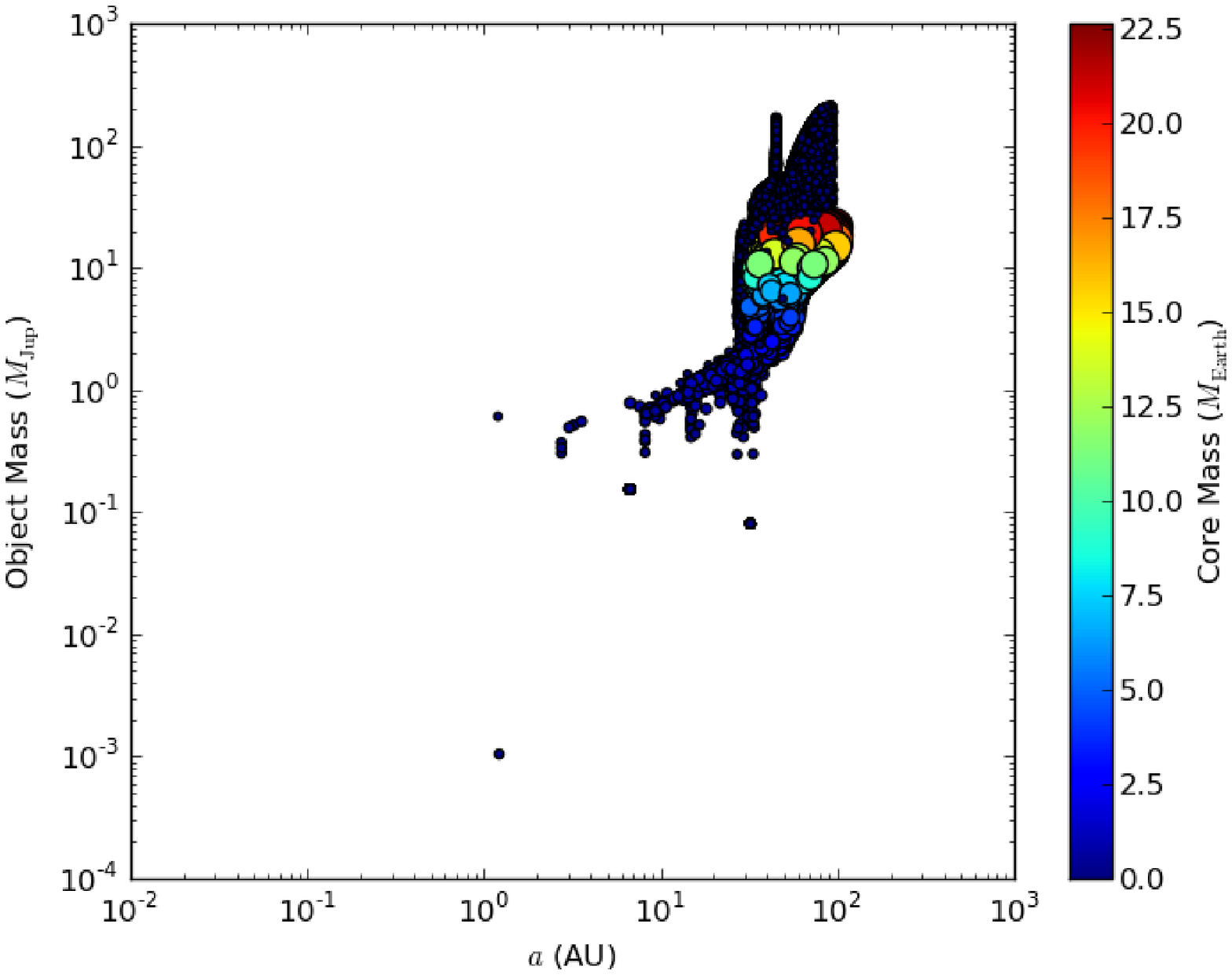} \\
\end{array}$
  \caption{Embryo mass versus semi-major axis for $p_{\kappa}=1$, with
    a disc that truncates at 50 AU after fragmentation and migration timescales
    increased by a factor of 10. The left hand panel shows the initial
    fragment mass-semimajor axis distribution, with the colour bar
    indicating the timescale on which grains will sediment once
    sufficiently grown.  202,385 disc fragments were produced in this
    run. The right hand panel shows the final distribution after
    $10^6$ years.  Of the fragments produced, only 92,295 survived the
    tidal downsizing process, and 40,571 were able to form solid
    cores. The colours in this plot indicate the mass of the core in
    Earth masses. \label{fig:m_vs_a_pkap1slowmig}}
\end{center}
\end{figure*}

\noindent In the same vein as \citet{Ida_Lin_mig}, we now vary the
speed of migration by multiplying the migration timescales in
equations (\ref{eq:typeI}) and (\ref{eq:typeII}) by a factor of 10.
Figure \ref{fig:m_vs_a_pkap1slowmig} shows the initial and final mass
versus semimajor axis distributions in this run.

With migration ten times less efficient than the previous run,
semi-major axes lower than the fragmentation boundary are less
populated than before.  Some objects still migrate to the disc inner
boundary, but in much fewer numbers than before.  The sedimentation
timescale for the fragments remains the same, so the cores produced
are very similar to the previous case.  With the ratio between
sedimentation and migration timescale being reduced by a factor of
ten, the formation of solid cores before disruption is somewhat
easier.  Indeed, this run shows the only example of a terrestrial
planet being formed in the simulation, with mass less than 1 $\mearth$
at a semimajor axis of 1 au. 

\begin{figure*}
\begin{center}$\begin{array}{cc}
\includegraphics[scale = 0.4]{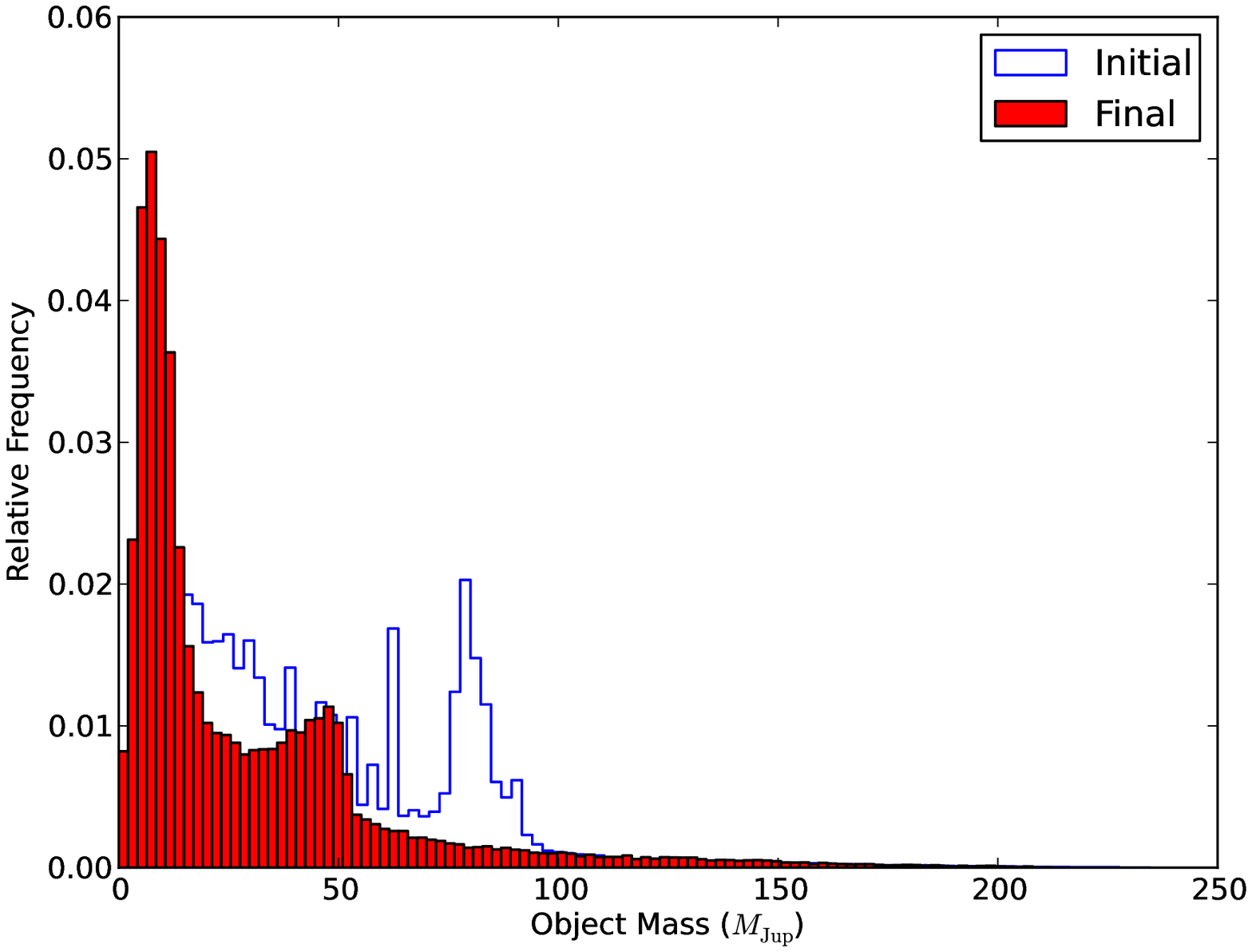} &
\includegraphics[scale = 0.4]{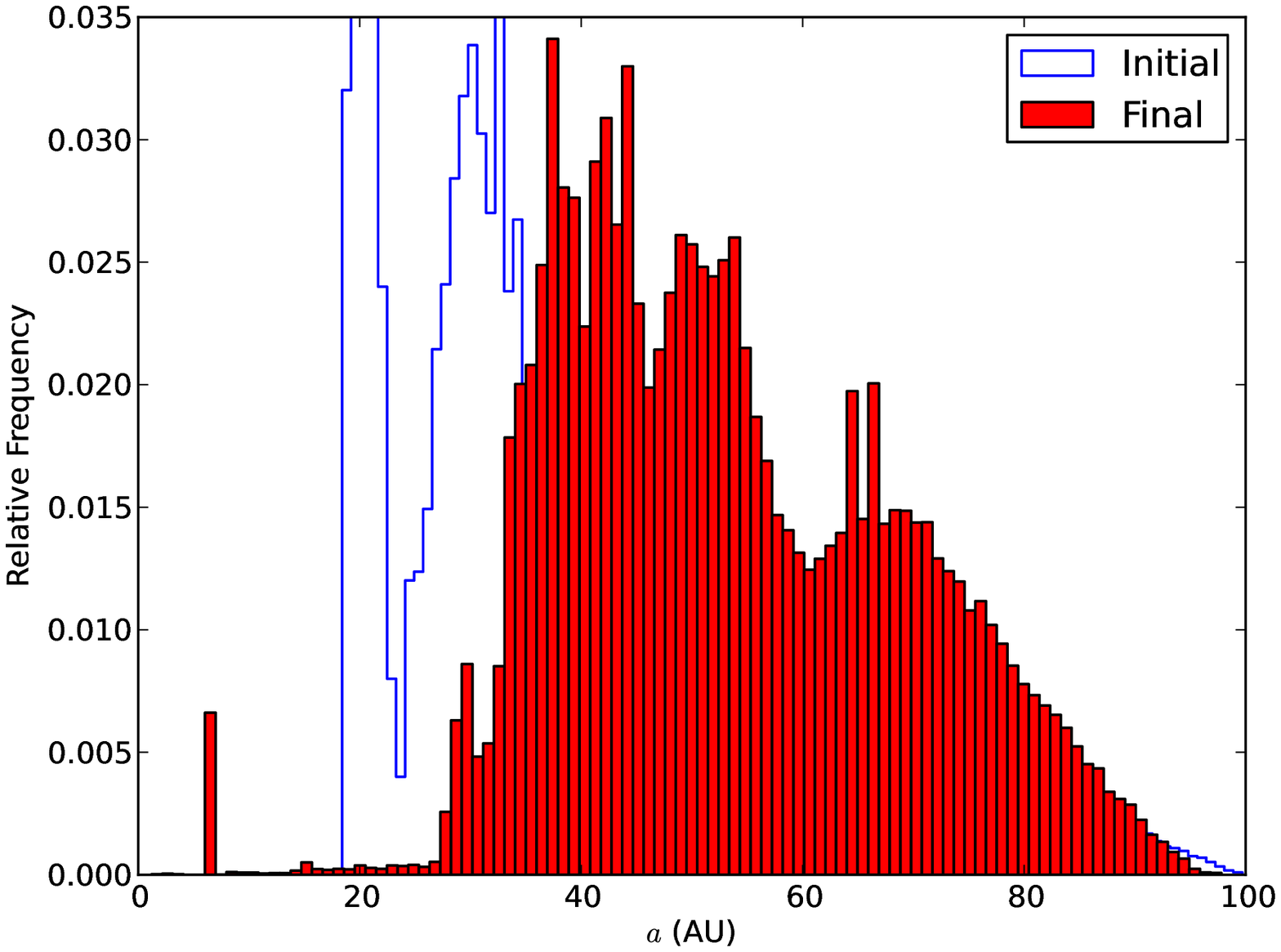} \\
\end{array}$
\caption{Comparing the initial (blue lines) and final (red bars)
  distributions of embryo mass (left) and embryo semimajor axis
  (right).\label{fig:compare_pkap1slowmig}}
\end{center}
\end{figure*}

The embryo mass distribution (Figure \ref{fig:compare_pkap1slowmig},
left panel) remains relatively unchanged, but the semimajor axis
distribution (right panel) shows an even larger desert, with few
objects ending up inside 20 au.

\begin{figure*}
\begin{center}$\begin{array}{cc}
\includegraphics[scale = 0.4]{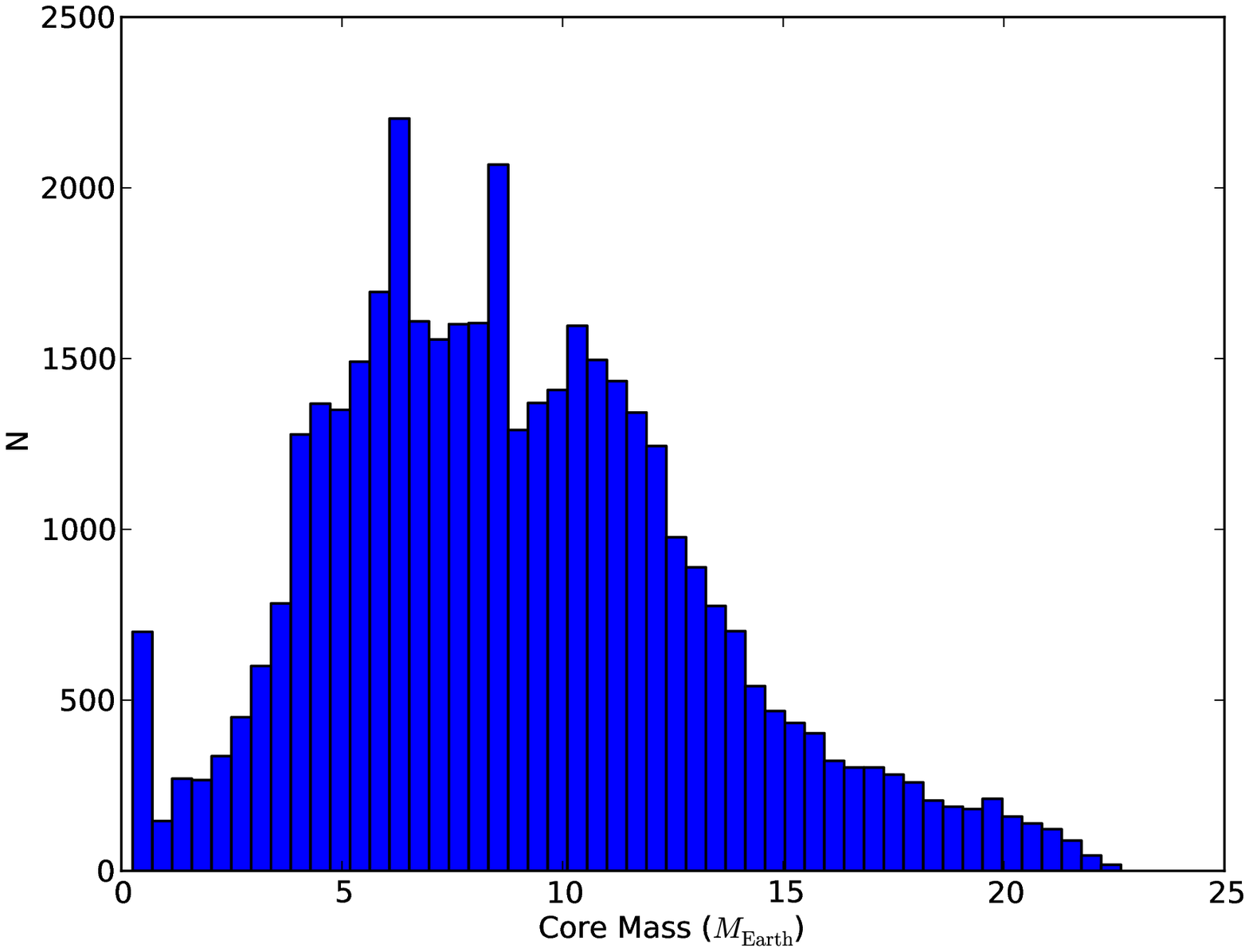} &
\includegraphics[scale = 0.4]{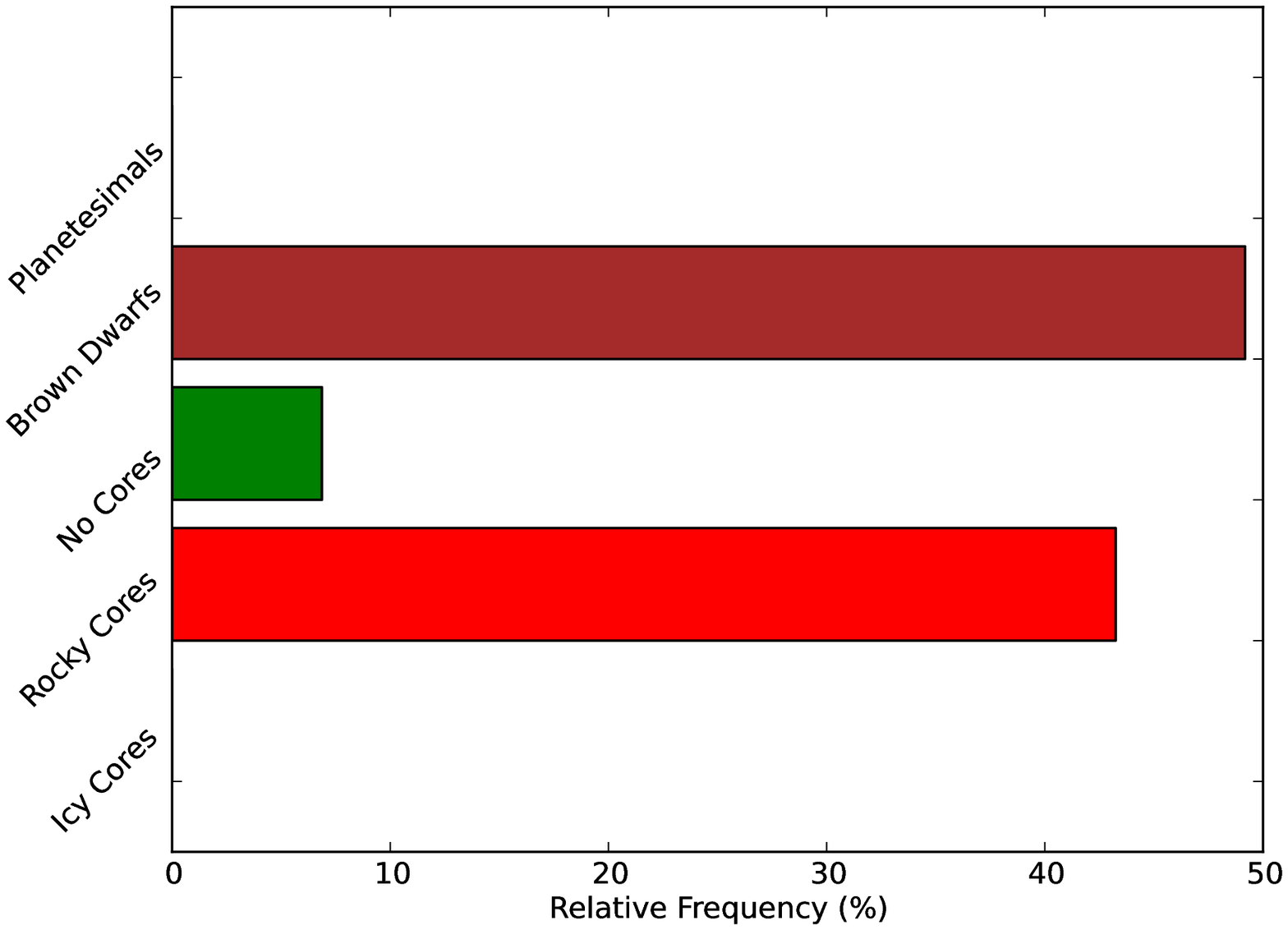} \\
\end{array}$
\caption{Left: The distribution of core masses after $10^6$
  years of evolution.  Right: The relative frequency of each object
  type formed in the population synthesis model.
  (right).\label{fig:mcore_types_pkap1slowmig}}
\end{center}
\end{figure*}

The core mass distribution (left panel of Figure
\ref{fig:mcore_types_pkap1slowmig}) is again very similar as the
sedimentation behaviour is unchanged.  Brown dwarfs still dominate the
population (right panel of Figure \ref{fig:mcore_types_pkap1slowmig}),
followed closely by cored objects.

\subsection{Opacity Law Index $p_{\rm \kappa} = 2$ (Truncated Disc)}

\begin{figure*}
\begin{center}$\begin{array}{cc}
\includegraphics[scale = 0.4]{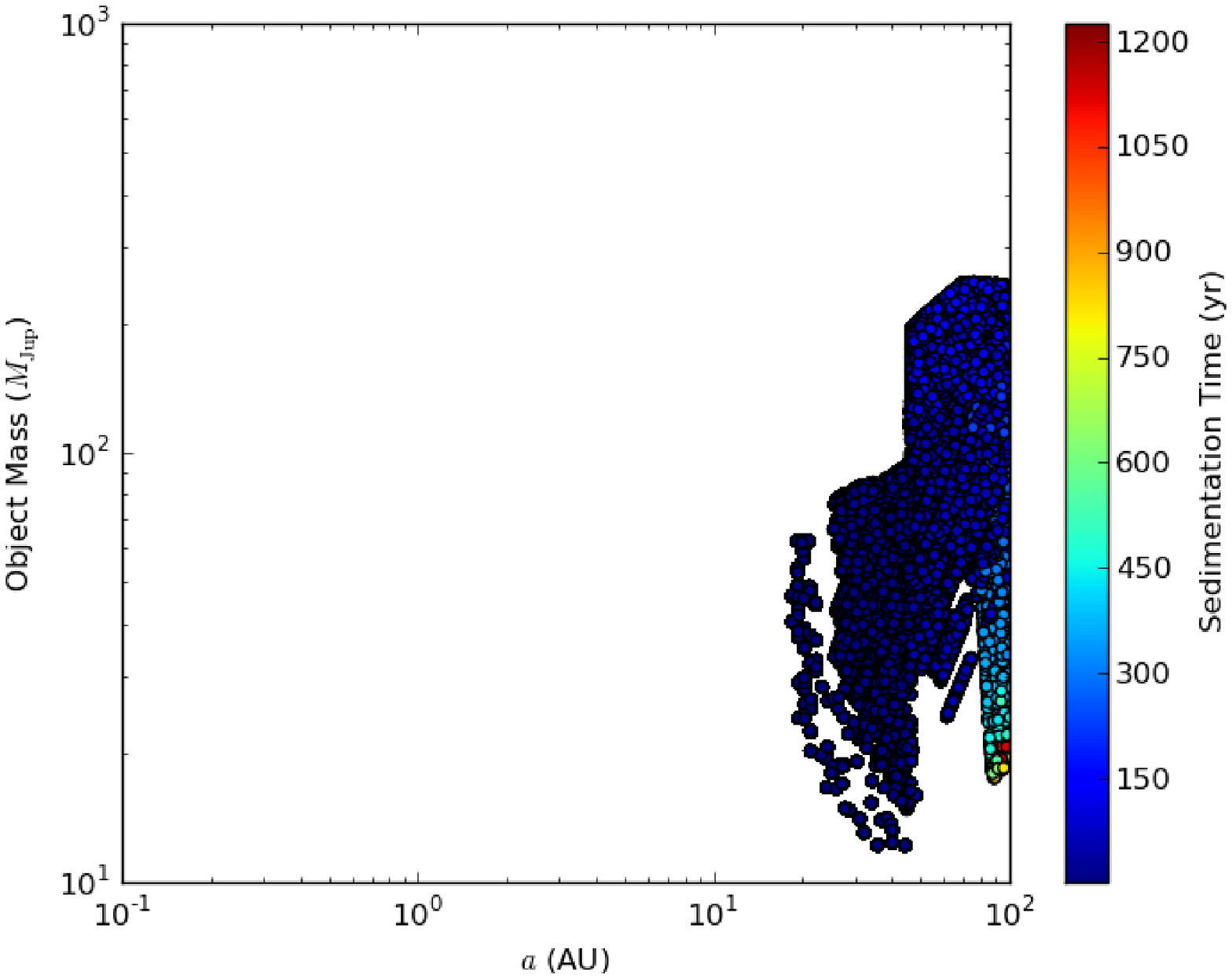} &
\includegraphics[scale=0.4]{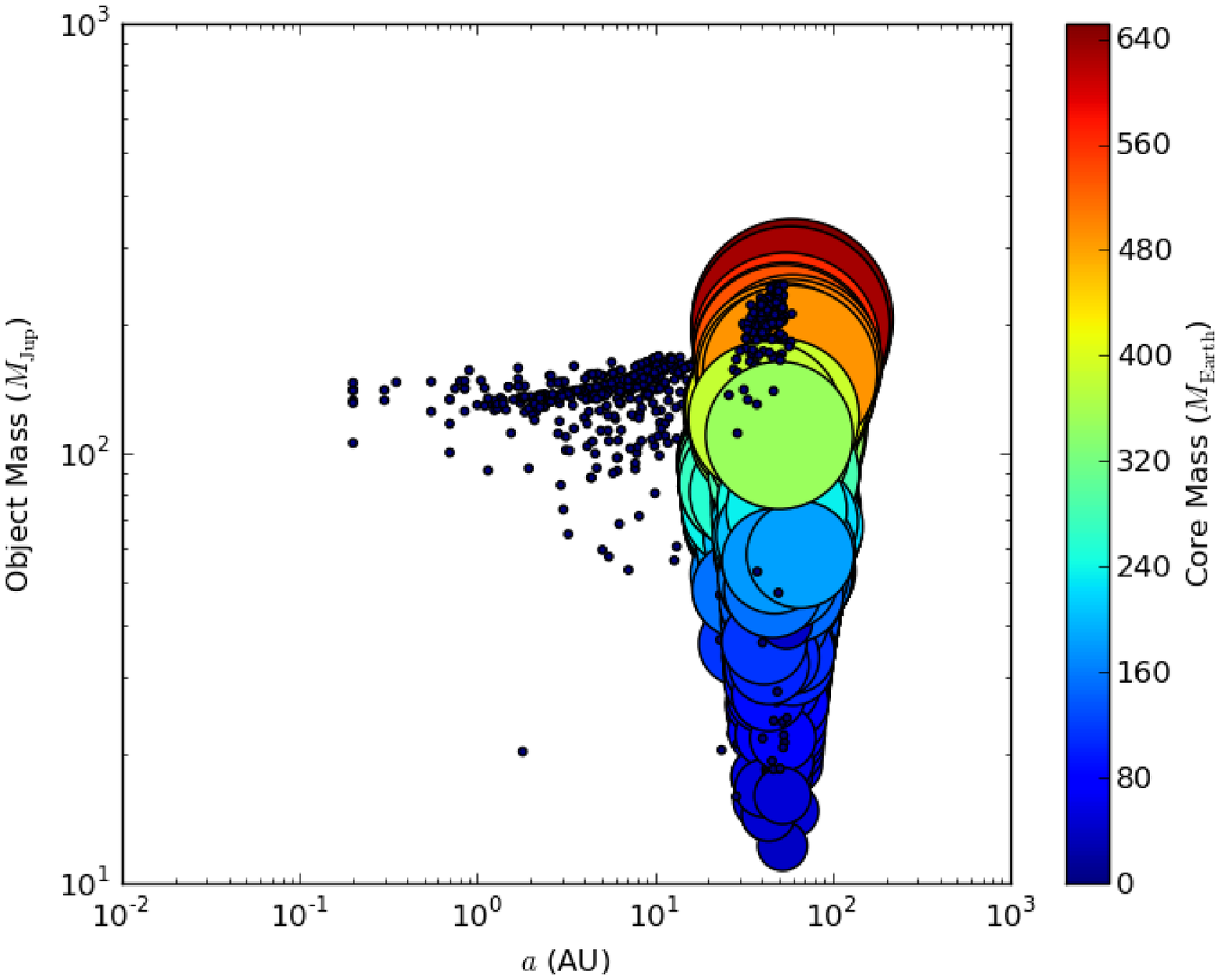} \\
\end{array}$
  \caption{Embryo mass versus semi-major axis for $p_{\kappa}=2$, with
    a disc that truncates at 50 AU after fragmentation. The left hand panel
    shows the initial fragment mass-semimajor axis distribution, with
    the colour bar indicating the timescale on which grains will
    sediment once sufficiently grown.  530,488 disc fragments were
    produced in this run. The right hand panel shows the final
    distribution after $10^6$ years.  Of the fragments
    produced, only 22,637 survived the tidal downsizing process, and
    2,292 were able to form solid cores. The colours in this plot
    indicate the mass of the core in Earth masses.  The left hand panel
    shows the initial fragment mass-semimajor axis distribution, with
    the colour bar indicating the timescale on which grains will
    sediment once sufficiently grown. \label{fig:m_vs_a_pkap2}}
\end{center}
\end{figure*}

\noindent In this run, we now allow migration to be efficient, and
truncate the disc at fragmentation, but change the opacity law so that
it scales with the square of the temperature.  This significantly
increases the cooling time of the embryo, which will have important
consequences for the population produced.  Figure
\ref{fig:m_vs_a_pkap2} shows the initial and final mass vs semimajor
axis distributions for this run.

\begin{figure*}
\begin{center}$\begin{array}{cc}
\includegraphics[scale = 0.4]{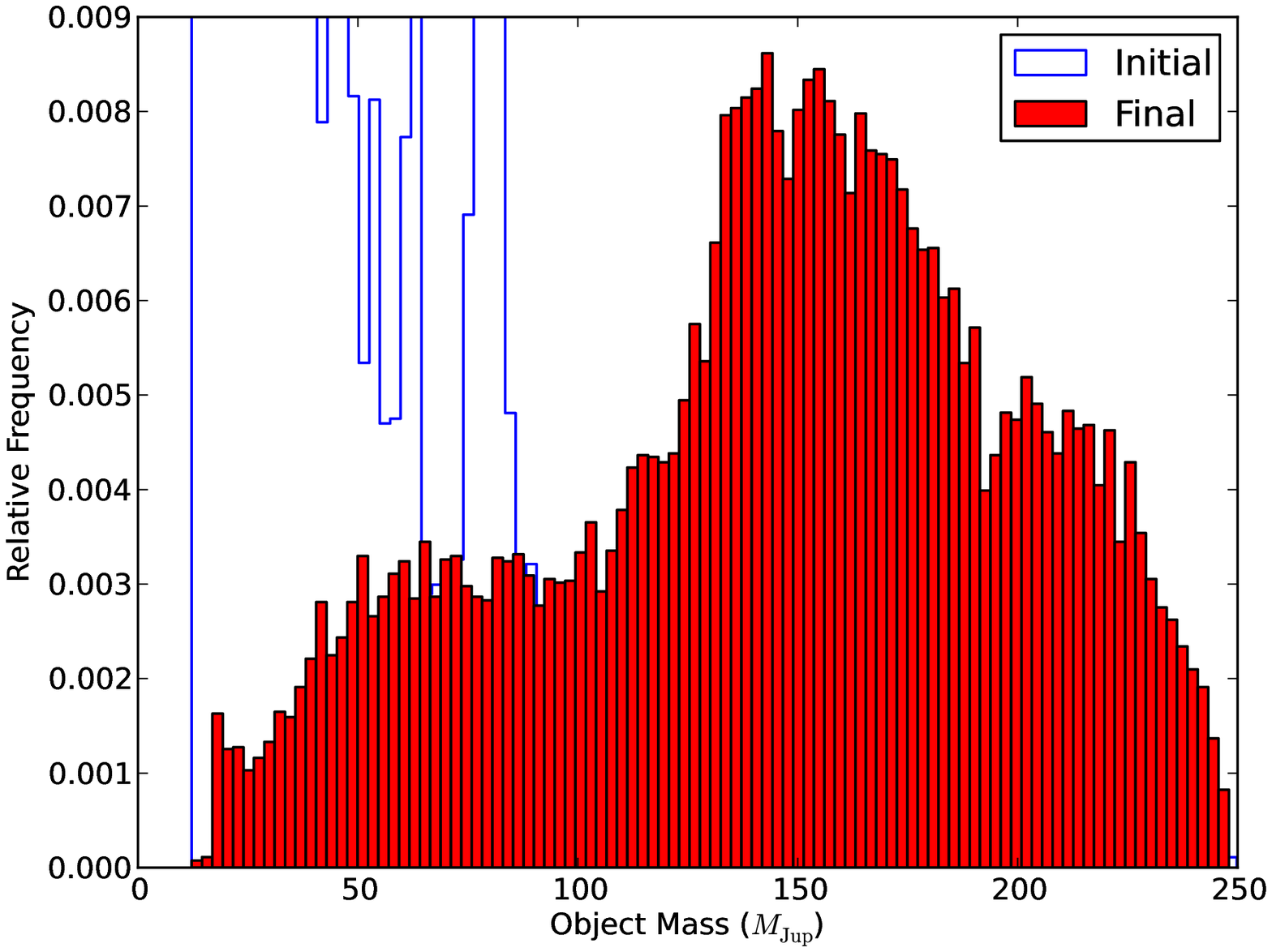} &
\includegraphics[scale = 0.4]{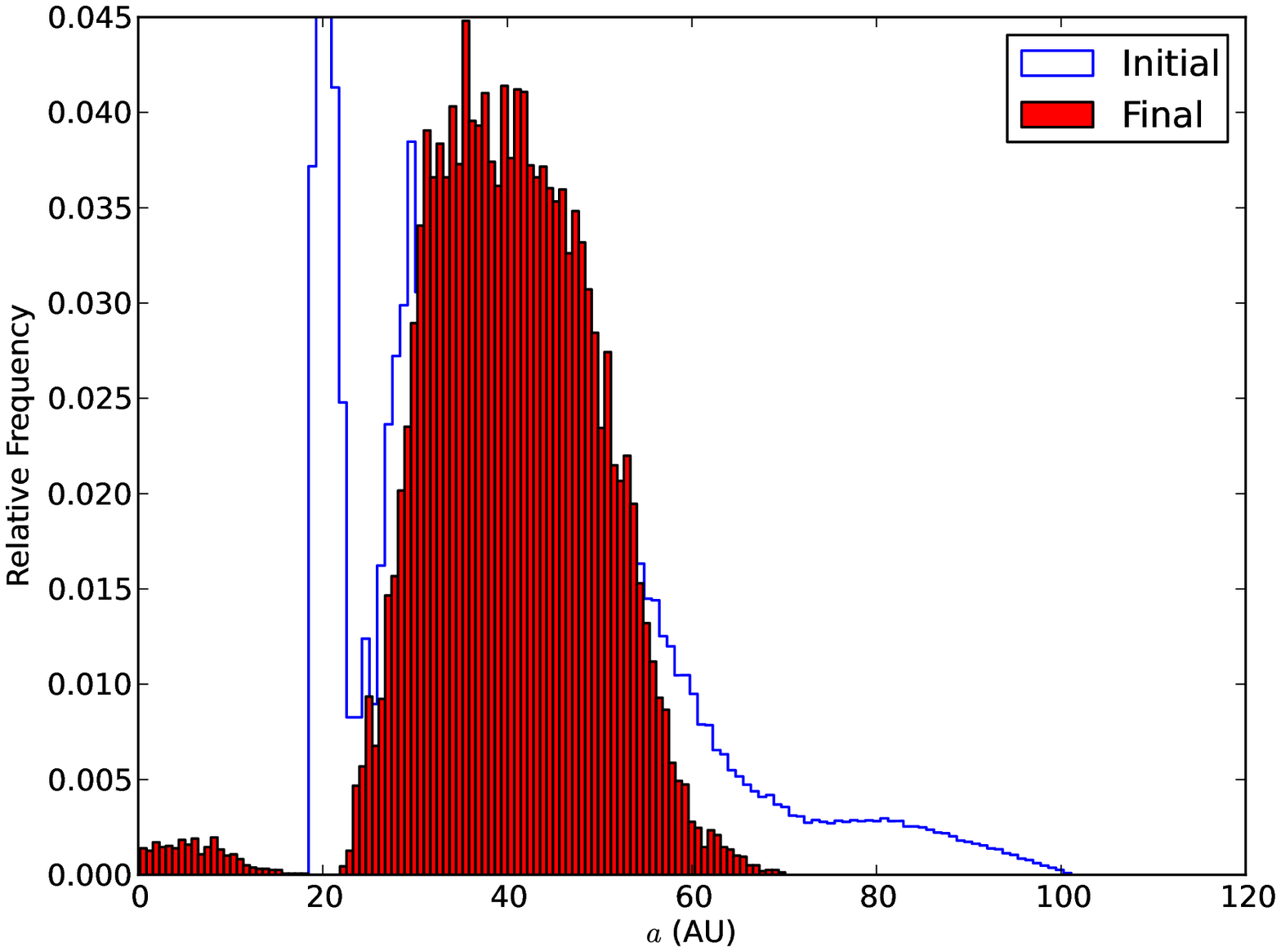} \\
\end{array}$
\caption{Comparing the initial (blue lines) and final (red bars)
  distributions of embryo mass (left) and embryo semimajor axis
  (right).\label{fig:compare_pkap2}}
\end{center}
\end{figure*}

The inefficient cooling of the embryo results in a greatly increased
destruction rate - over 95\% of the fragments produced in this run are
destroyed.  Only the most massive objects survive, and there are only
a handful that survive in the inner disc.  This is reflected in the
mass and semimajor axis distributions (Figure
\ref{fig:compare_pkap2}).

\begin{figure*}
\begin{center}$\begin{array}{cc}
\includegraphics[scale = 0.4]{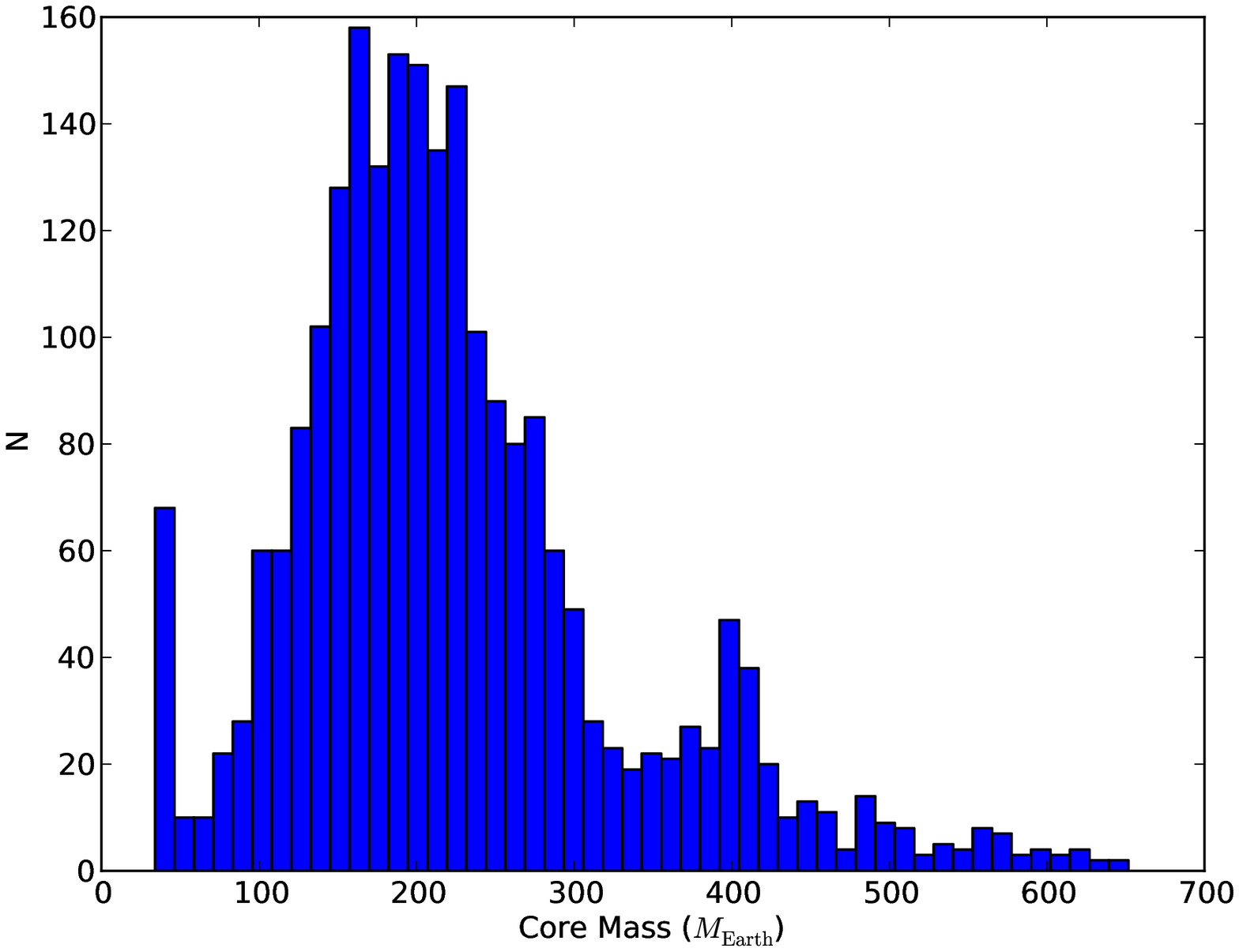} &
\includegraphics[scale = 0.4]{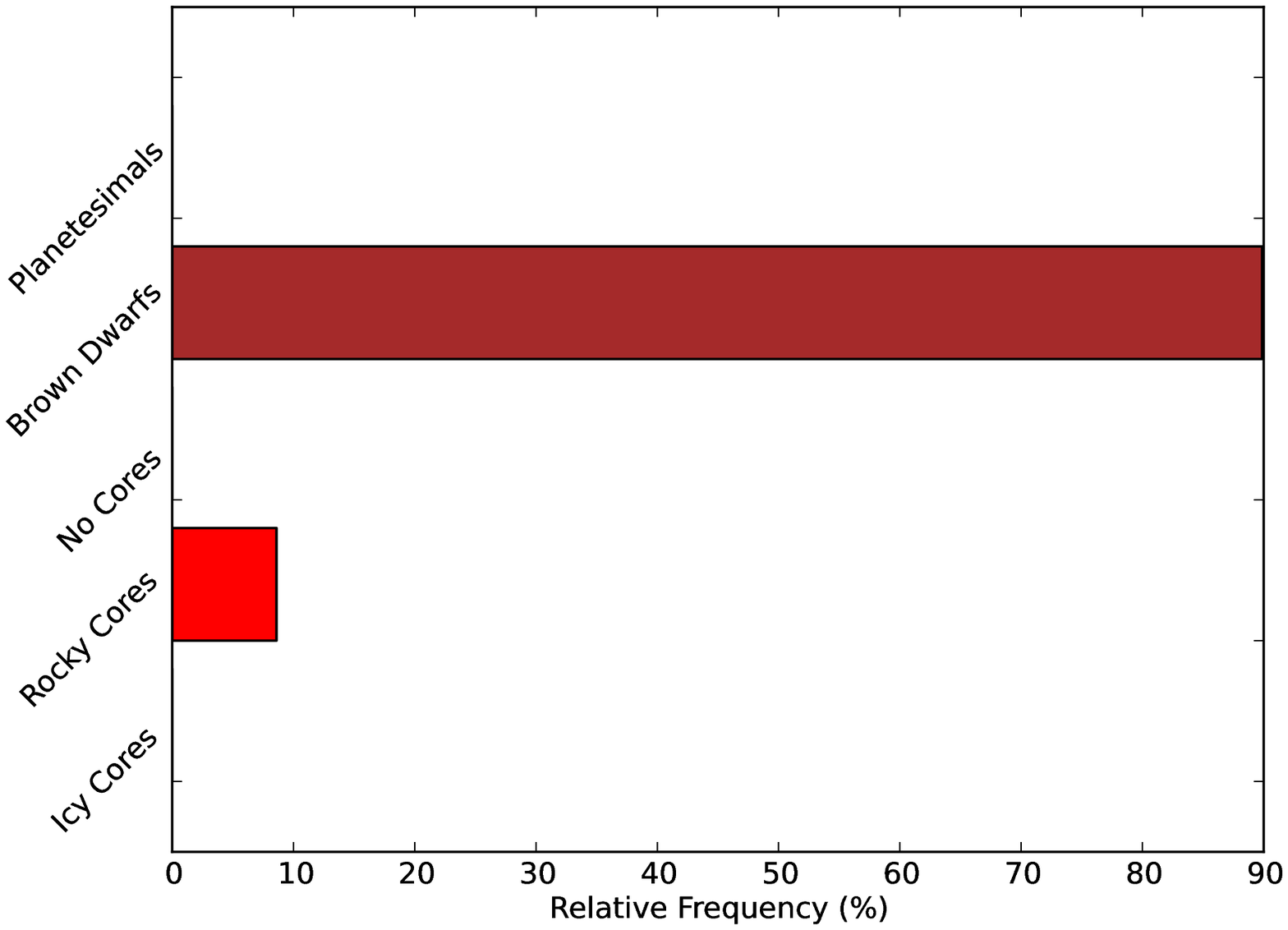} \\
\end{array}$
\caption{Left: The distribution of core masses after $10^6$
  years of evolution.  Right: The relative frequency of each object
  type formed in the population synthesis model.
  (right).\label{fig:mcore_types_pkap2}}
\end{center}
\end{figure*}

Perhaps most worrying are the extremely large solid core masses
produced in this run (left panel of Figure
\ref{fig:mcore_types_pkap2}).  The mode of this distribution is around
$200 \mearth \sim 0.5 \mjup$.  Note that these very large cores only
exist in massive gaseous embryos - there are no terrestrial type
objects (i.e. solid cores without gaseous envelopes).  These
anomalously large cores are suggestive of missing physics from the
model (see Discussion).

The inefficiency of embryo cooling only allows massive objects to
remain bound without becoming disrupted, and as a result cores form
very rarely in this run (only around 10\% of all surviving objects).
The other 90\% form brown dwarfs - indeed, some are massive enough to
form hydrogen-burning low-mass stars with masses $\sim 0.1 \msol$.

\subsection{Opacity Law Index $p_{\rm \kappa} = 1$ (No Disc Truncation)}  

\noindent We have seen in the previous runs that the migration of the
embryos strongly determines their mass and semimajor axis evolution.
Until now, we have assumed that the fragmentation process truncates
the disc, which allows embryos to evolve their physical radius and
dust population before the disc can undergo viscous spreading,
extending its outer radius until it encroaches upon their orbital
domains, and begin the migration process.

In this run, we do not truncate the disc at fragmentation.  The disc
can therefore begin migrating the embryos immediately, decreasing the
Hill Radius of the embryos precipitately and significantly increasing
the rate of disruption.  As the $p_{\rm \kappa}=2$ run already shows a
high destruction rate, we reset $p_{\rm \kappa}$ to 1 to assess the
significance of disc truncation.

\begin{figure*}
\begin{center}$\begin{array}{cc}
\includegraphics[scale = 0.4]{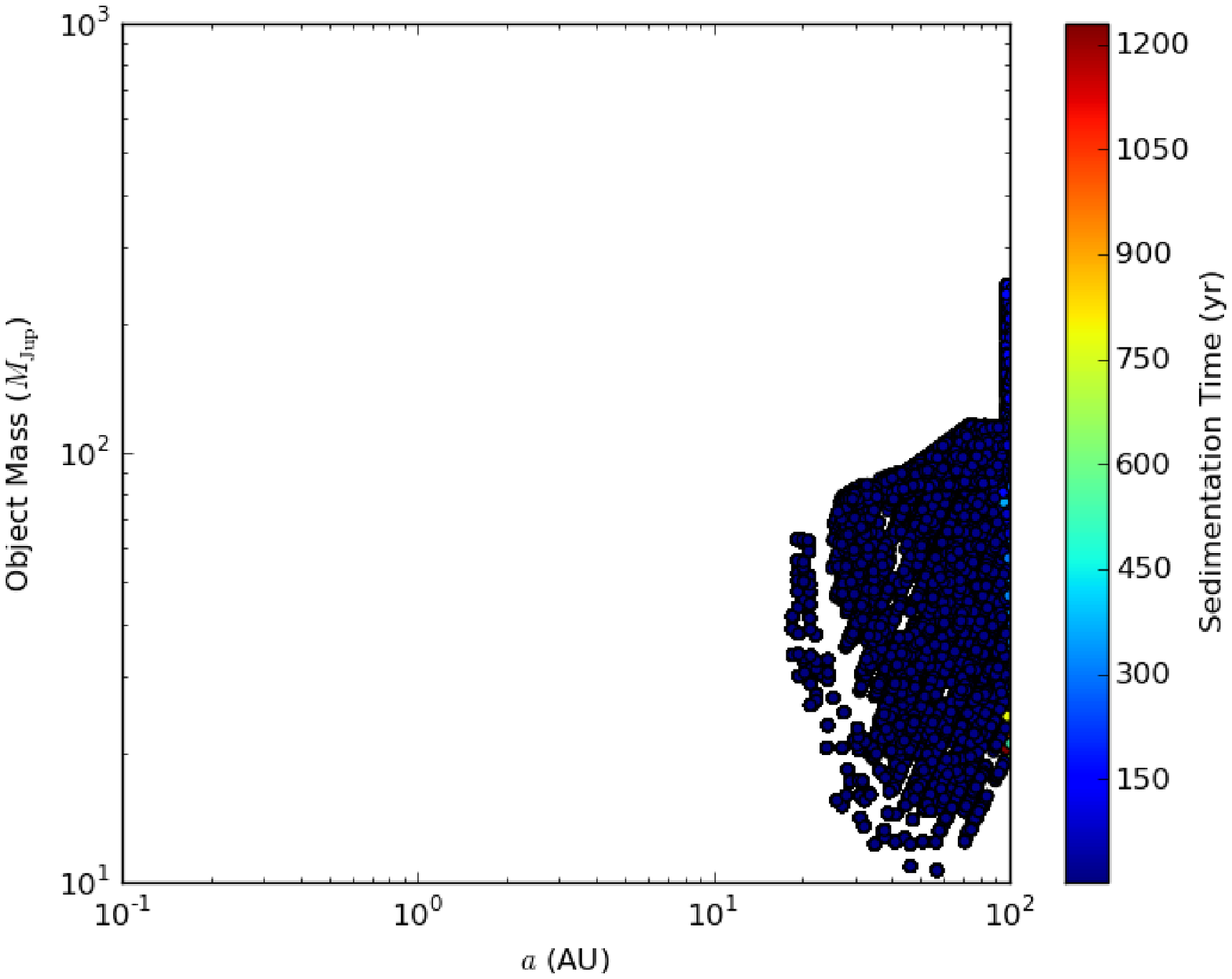} &
\includegraphics[scale=0.4]{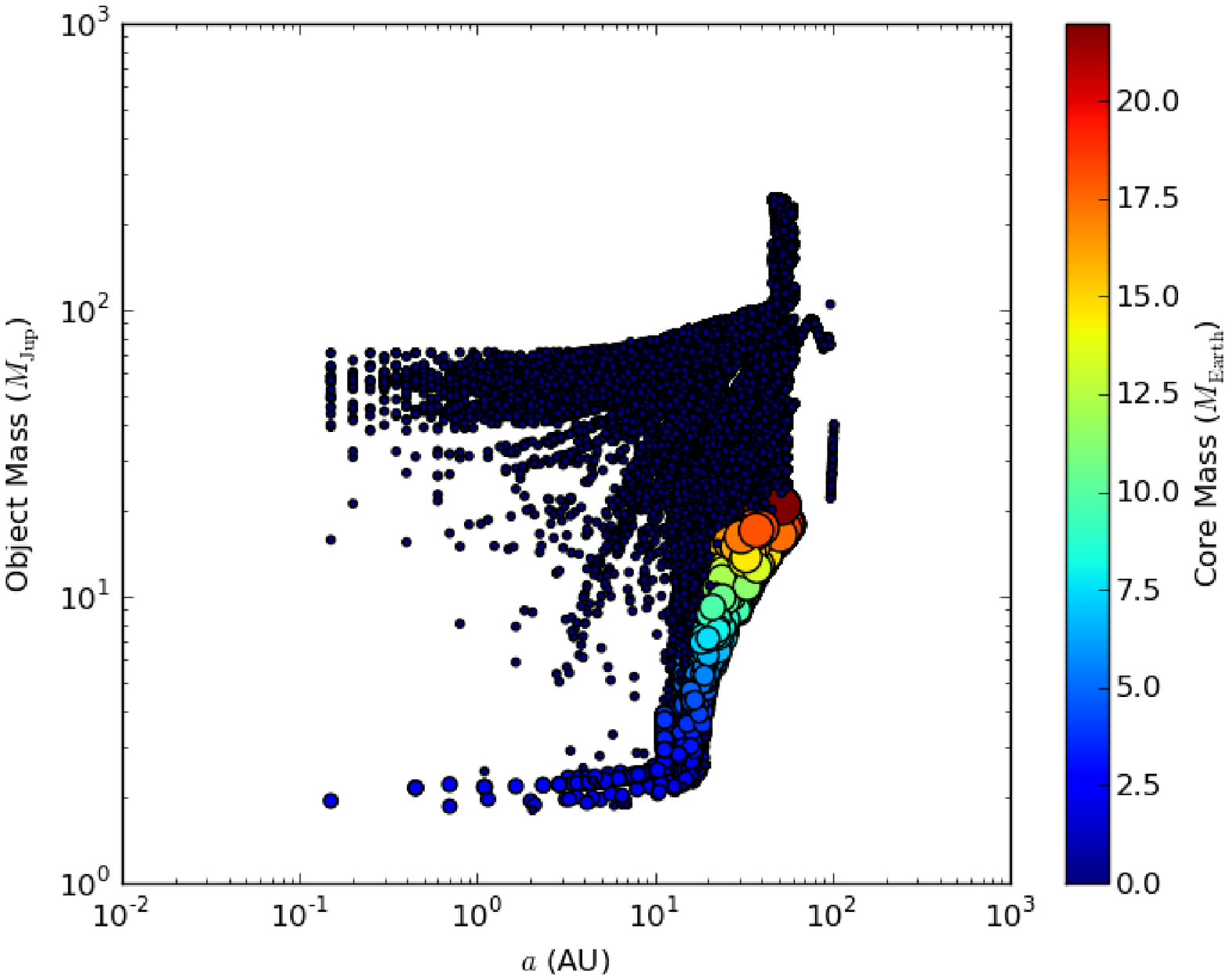} \\
\end{array}$
  \caption{ Embryo mass versus semi-major axis for $p_{\kappa}=1$, with
    a disc that does not truncate after fragmentation. The left hand panel
    shows the initial fragment mass-semimajor axis distribution, with
    the colour bar indicating the timescale on which grains will
    sediment once sufficiently grown.  205,184 disc fragments were
    produced in this run. The right hand panel shows the final
    distribution after $10^6$ years.  Of the fragments
    produced, only 74,430 survived the tidal downsizing process, and
    7,378 were able to form solid cores. The colours in this plot
    indicate the mass of the core in Earth masses.  The left hand panel
    shows the initial fragment mass-semimajor axis distribution, with
    the colour bar indicating the timescale on which grains will
    sediment once sufficiently grown. \label{fig:m_vs_a_pkap1notrunc}}
\end{center}
\end{figure*}

Figure \ref{fig:m_vs_a_pkap1notrunc} shows the initial and final mass
vs semimajor axis distributions, which indicate that removing disc
truncation has a significant effect.  Again, there are no terrestrial
type objects, and all objects which possess a core possess a massive
envelope.  With migration occuring immediately after fragmentation,
many more objects reach the inner disc boundary, producing a small
population of Hot Jupiters and close BD companions.  Core formation is
suppressed compared to the other $p_{\kappa}=1$ runs: cores form in
10\% of surviving embryos compared to around 40\% in the other $p_{\rm
  \kappa}$ cases.

\begin{figure*}
\begin{center}$\begin{array}{cc}
\includegraphics[scale = 0.4]{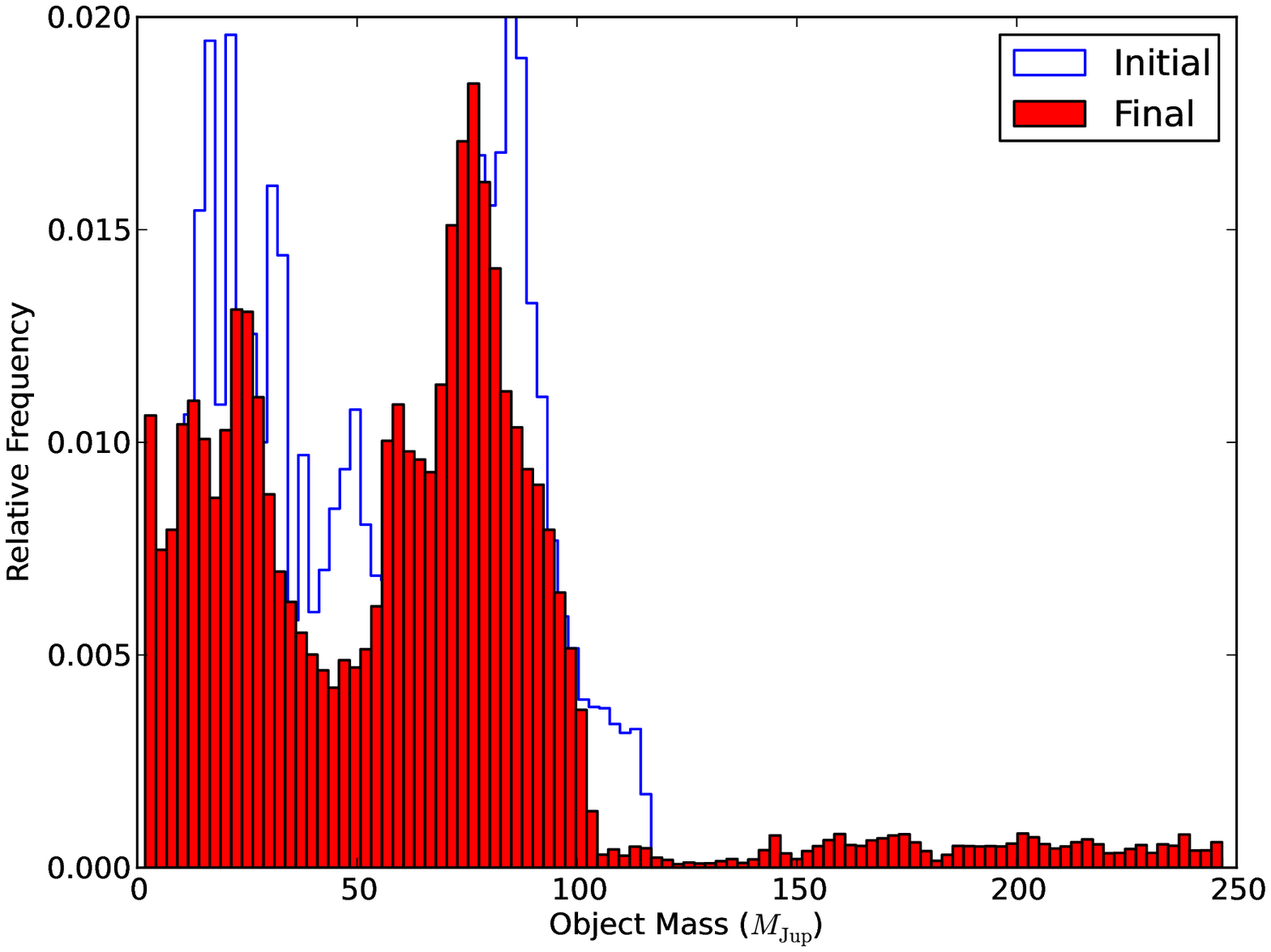} &
\includegraphics[scale = 0.4]{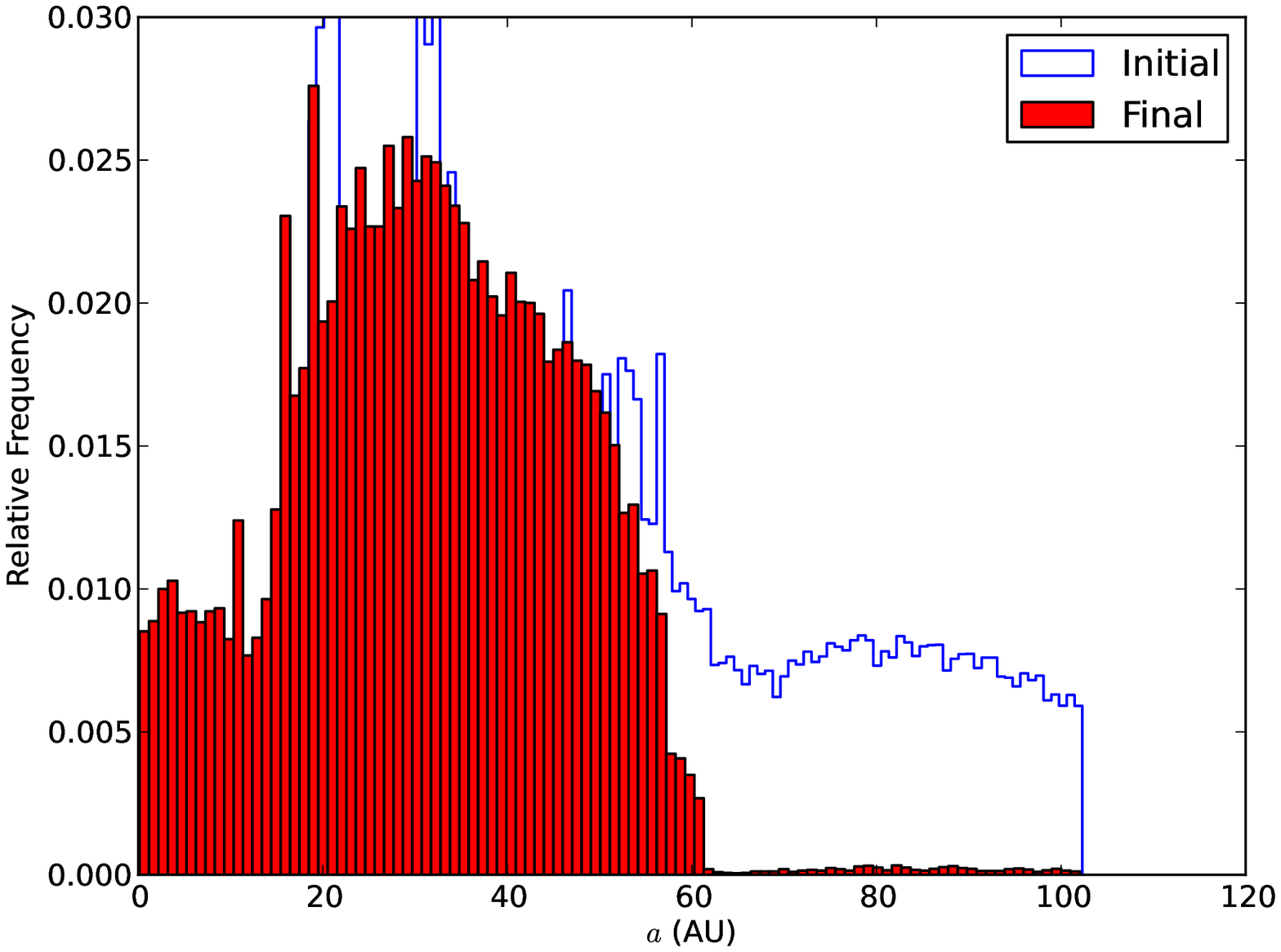} \\
\end{array}$
\caption{Comparing the initial (blue lines) and final (red bars)
  distributions of embryo mass (left) and embryo semimajor axis
  (right).\label{fig:compare_pkap1notrunc}}
\end{center}
\end{figure*}

The mass distribution of the objects (left panel of Figure
\ref{fig:compare_pkap1notrunc}) remains relatively unchanged from
initial to final, forming large numbers of objects up to 100 $\mjup$.
The two modes of the distributions are generated by the spacing
of the fragments as they form in the disc (with some broadening of the
peaks due to the uniform sampling of the spacing in Hill Radii
described in earlier sections).  

This population is very much a stellar population.  The brown dwarf
desert is less significant (right panel of Figure
\ref{fig:compare_pkap1notrunc}), and few objects survive with semimajor
axis larger than 60 au.

\begin{figure*}
\begin{center}$\begin{array}{cc}
\includegraphics[scale = 0.4]{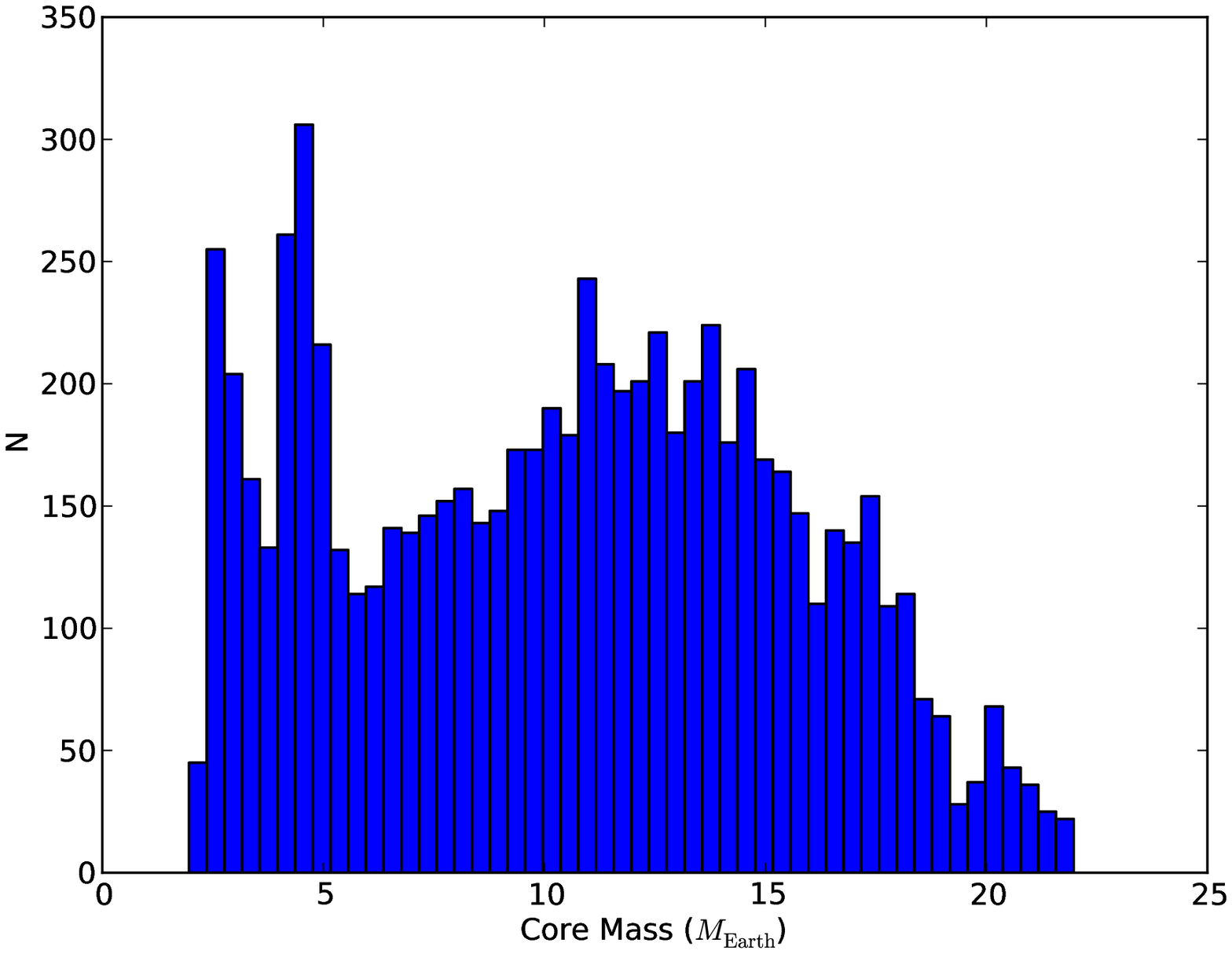} &
\includegraphics[scale = 0.4]{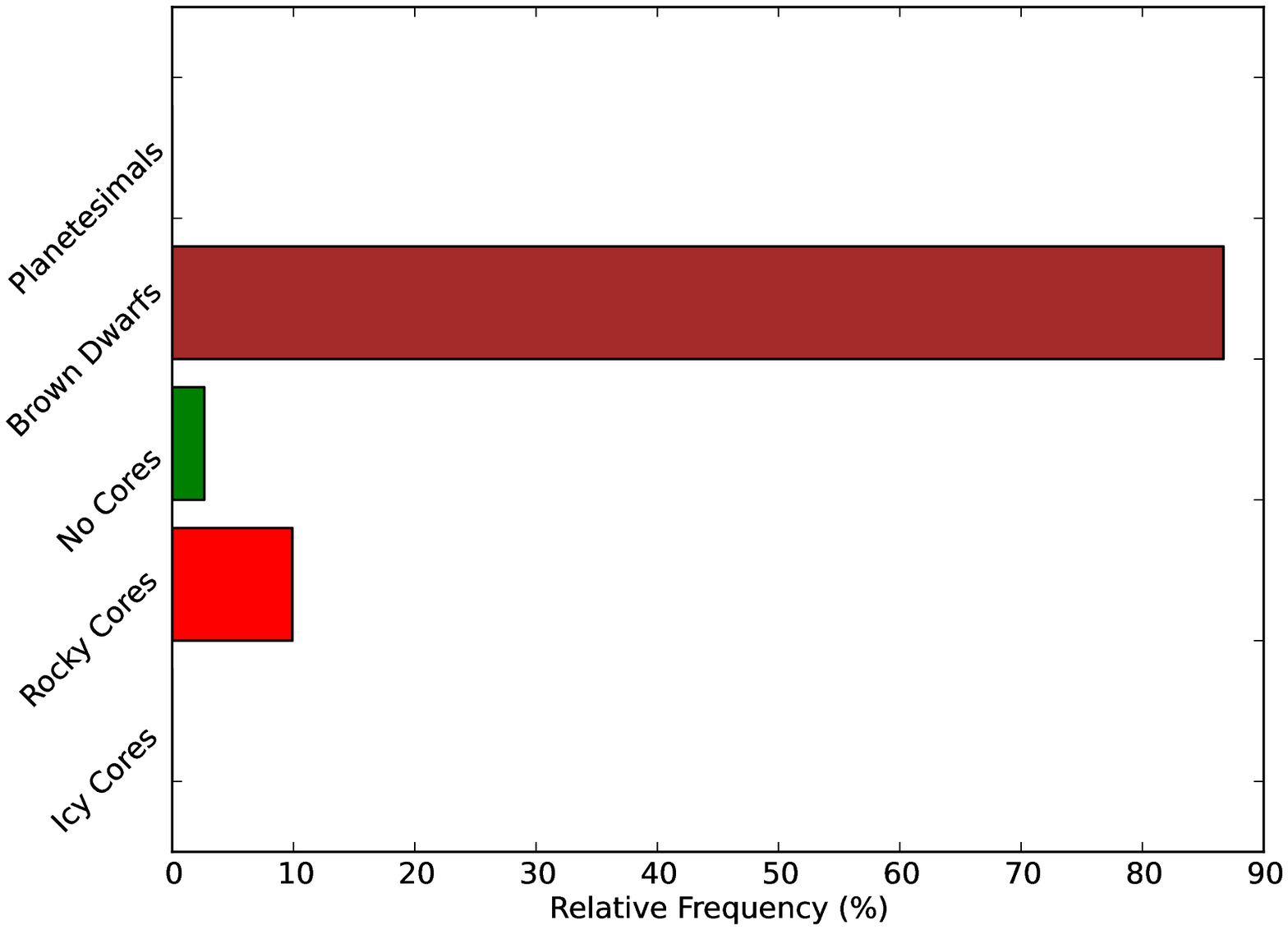} \\
\end{array}$
\caption{Left: The distribution of core masses after $10^6$
  years of evolution.  Right: The relative frequency of each object
  type formed in the population synthesis model.
  (right).\label{fig:mcore_types_pkap1notrunc}}
\end{center}
\end{figure*}

The core mass distribution (left panel of Figure
\ref{fig:mcore_types_pkap1notrunc}) also reflects the initial mass
distribution. Less than 10\% of all surviving objects have cores
(right panel of Figure \ref{fig:mcore_types_pkap1notrunc}), and nearly
90\% of all the objects are brown dwarfs (or low mass stars).

\section{Discussion}\label{sec:discussion}

\subsection{Limitations of the Model}

\noindent We readily admit that this model, while detailed in
construction, still lacks some important physics, which we hope to
include in future work.

\subsubsection{Accretion of Mass and Angular Momentum by the Embryo}

\noindent Perhaps the most glaring omission from this population
synthesis model is accretion of material onto the embryos after
fragmentation.  As the semi analytic disc model is evolved without the
presence of fragments, modelling subsequent mass loss from the disc
onto embryos is not possible.  The embryo should be able to accrete
provided it does not open a significant gap, which it will do if a)
its Hill radius exceeds the local disc scale height, and b) the clump
is sufficiently massive to overcome the local Reynolds turbulence
\citep{Lin1979} .  For almost all fragments formed, this criterion is
easily satisfied initially.  However, as the embryo migrates inward,
the Hill radius decreases, and the embryo's mass also decreases due to
tidal disruption.  The gap opening criterion will no longer be
satisfied, and accretion can begin once more.

To what level this latter accretion phase will affect the subsequent
evolution of fragments is unclear.  Numerical simulations by
\citet{Zhu2012} show gap opening at semimajor axes as low as 10 au,
which would be a consequence of accretion increasing the Hill radius
past the value of the local scale height.

We have also ignored the embryo's ability to accrete solids from its surroundings \citep{Helled2006}.  The embryo may be able to substantially enrich itself during the early stages of its contraction, accreting a large fraction of the available planetesimals in its feeding zone, boosting the dust-to-gas ratio.  This could be incorporated into our model quite simply using simple two-body capture approximations.

The accretion history of the embryo will determine its angular
momentum.  \citet{Cha2011} find in their numerical simulations that
the fragments produced rotate prograde to the disc's rotation, with a
rotational angular frequency approximately 10\% of their maximum
break-up frequency.  This can be used to estimate the rotational
angular frequency of terrestrial planets formed from these embryos
whose maximum exceeds the break-up frequency \citep{Nayakshin2011a}.
This would suggest that embryos rotate sufficiently rapidly to form circum-embryonic discs \citep{Boley2010b,Shabram2013} and 
binary cores, an outcome not producible in this current model.

Equally, the rotation rate of giant planets is expected, as a rule, to
be initially prograde, and close to the break-up frequency, although
stellar tides can damp this rotation rate.  Also, events such as
planet-planet scattering or the Kozai mechanism can produce retrograde
orbits (e.g. \citealt{Nagasawa2011}).  Future work should model this
angular momentum evolution in detail.

\subsubsection{Dust Coagulation}

The extremely massive cores produced by the $p_{\kappa}=2$ run suggest
that there is missing physics in the coagulation of the dust grains.
While turbulence is part of the coagulation process in these models,
convection is not.  As we have mentioned previously, allowing the envelope to be convective can
significantly increase the sedimentation timescale
\citep{Helled2008a}, although it does not completely preclude core
formation, especially for low-mass embryos \citep{Helled2008}.  Also, we should strictly alter the opacity in the embryo
as the grains sediment, as the local metallicity of the gas can change
significantly.  Global metallicity effects are also important:
decreasing the initial dust to gas ratio in the disc can greatly decrease the
contraction timescale of the embryo \citep{Helled2011}.

\subsubsection{Embryo Migration}

We have deliberately implemented a basic model of planetary migration
in this work, as this allows us to investigate the complete system of
equations presented by
\citet{Nayakshin2010a,Nayakshin2010b,Nayakshin2011} and assess their
expected output on a statistical level.  As such, the modelling of
migration misses out important characteristics of the migration
process: stochasticity due to disc turbulence \citep{Baruteau2011},
resonant migration of multiple bodies (e.g. \citealt{Libert2011}), and
halting of migration at planet traps induced by disc structure (e.g
\citealt{Kretke2012}).  Inwardly migrating giant planets can even halt
or reverse the motion of low mass planets at larger semimajor axis
\citep{Podlewska-Gaca2012}.

Also, the eccentricity and inclination evolution of the bodies is not
tracked: recent numerical simulations suggest that eccentricity
excitation via planet-disc interaction will be low for planet masses
of a few Jupiter masses \citep{Dunhill2012}, which appears to be borne
out by observations \citep{Dawson2012}, but future models should
investigate this by allowing eccentricities to be excited and damped.

Most migration models are developed for low-mass, non-self-gravitating
discs: simulations of fragment migration in self-gravitating discs
have shown that the usual embryo migration equations do not appear to
apply, and that fragment migration is typically much faster
\citep{Michael2011,Baruteau2011,Zhu2012}.  Also, as our embryos are evolved
independently of each other, we cannot model for example the chaotic
orbital interchanges witnessed in self-gravitating disc simulations
with multiple embryos \citep{Boss2013}.  It is unclear how these interactions can be modelled semi-analytically, but it is clear that any model output should be considered with the absence of this physical process at the forefront of one's mind.

Can these simulations produce semi-analytic prescriptions for
migration? \citet{Baruteau2011} show that gravitoturbulence reduces
the Type I migration timescale by a factor of a few as a result of the
disc's non-axisymmetry pushing the location of Lindblad resonances
towards the planet.  \citet{Zhu2012} also show that the Type I
timescale is altered by a factor which they fit in terms of embryo
mass and local surface density, accounting for the embryo's
inertia.  Future modelling should consider implementing some of these
initial empirical results.

\subsection{Terrestrial Planet Formation via Tidal Downsizing}

\noindent Despite the above limitations, we can make some reasonably
robust statements regarding the likelihood of terrestrial planet
formation.  All the runs carried out in this paper show that planets
of mass $\sim \mearth$ are extremely rare - out of over a million disc
fragments, only one stands out as a clear example of a terrestrial planet. 

Most of the missing physics described above is likely to inhibit the
formation of low mass cores rather than encourage it, so future
population synthesis modelling is likely to produce even fewer
terrestrial planets than we see here.  In short, it appears to be
clear that while GI followed by tidal downsizing can produce
terrestrial planets on some occasions, it is not an efficient means of
constructing such objects.  This would indicate that terrestrial
planets are much more likely to be formed by core accretion than disc
fragmentation.

\subsection{Production of Asteroid Belts}

\noindent For the production of asteroid belts via tidal downsizing,
the grain growth process must be well advanced, without the
sedimentation process forming a solid core or the embryo temperature
increasing above the vapourisation temperature.  If the disruption
process is sufficiently rapid, then the embryo and core will unbind,
and the grains in the embryo will be distributed in a ring at the
semimajor axis where disruption began \citep{Nayakshin2012}.

In these models, we did not see any embryos which completely satisfy
the above conditions for belt formation.  Only one run (the
non-truncated $p_{\kappa}=1$ run) showed some embryos with grains
surviving the disruption process without either being vapourised or
forming a core.  The embryos showed evidence of ice melting, but their
grains had not grown to the point where sedimentation would begin.  As
such, these would not form asteroid belts in the conventional sense,
but could perhaps form dust belts with some grains already grown (cf \citealt{Boley2010b}).

This would skew the disc grain size distribution towards larger
values, potentially providing a ``shortcut'' to grain growth in the
core accretion model.  This enhanced grain growth is similar to the
grain growth observed in non-fragmenting self-gravitating disc
spiral structures, which has also been proposed as a means of accelerating
the core accretion process \citep{Rice2004,Gibbons2012}. The key
difference between embryo-enhanced grain populations and spiral-wave
enhanced populations is the location of the grains' deposition:
spiral-wave enhanced grains will occupy semi major axes above several
tens of au \citep{Clarke2009}, whereas embryo-enhanced grains can be
delivered to lower semi major axis before the embryo is destroyed.

If sedimentation has just begun when the embryo is totally disrupted,
then a large fraction of the grains supplied to the disc may be of
order millimetre sizes, allowing rapid core accretion
\citep{Lambrechts2012}.

\subsection{Embryo disruption and Protostellar Outburst Models}

\noindent In this work, typically around 40\% of all embryos were
completely destroyed by tidal disruption.  This large destruction
rate, borne out by numerical simulations (e.g. \citealt{Boley2010,
  Nayakshin2010,Nayakshin2010a,Nayakshin2010b,Zhu2012}), could result
in protostellar outburst behaviour such as the FU Orionis phenomenon
\citep{Vorobyov_Basu_05,Vorobyov_Basu_06,Dunham2012}.  If this is the
case, then the behaviour of the gas as it becomes unbound from the
embryo will determine the nature of the burst.  Indeed, subsequent
disc fragmentation can be suppressed by the star's outburst
\citep{Stamatellos2012}.  We have not coupled disc evolution and
embryo evolution, and so we cannot model such behaviour.  As a result,
the population synthesis model does not characterise the outbursts,
but merely suggests that episodic stellar accretion is a common
outcome of the disc fragmentation process.

More subtly, \citet{Nayakshin2012a} show how a tidally disrupting
embryo can create deep gaps in the inner disc that are refilled by its
own matter, which can then be accreted onto the central star,
producing outbursts and quiescent periods as the gap refills, without
the embryo's destruction (see also \citealt{Lodato_Clarke_04}).  While
this population synthesis model does not self-consistently evolve the
embryo evolution and the disc evolution, this phenomenon could be
modelled semi-analytically in the future.

\subsection{Brown Dwarfs and the Desert}

\begin{figure*}
\begin{center}$\begin{array}{cc}
\includegraphics[scale = 0.4]{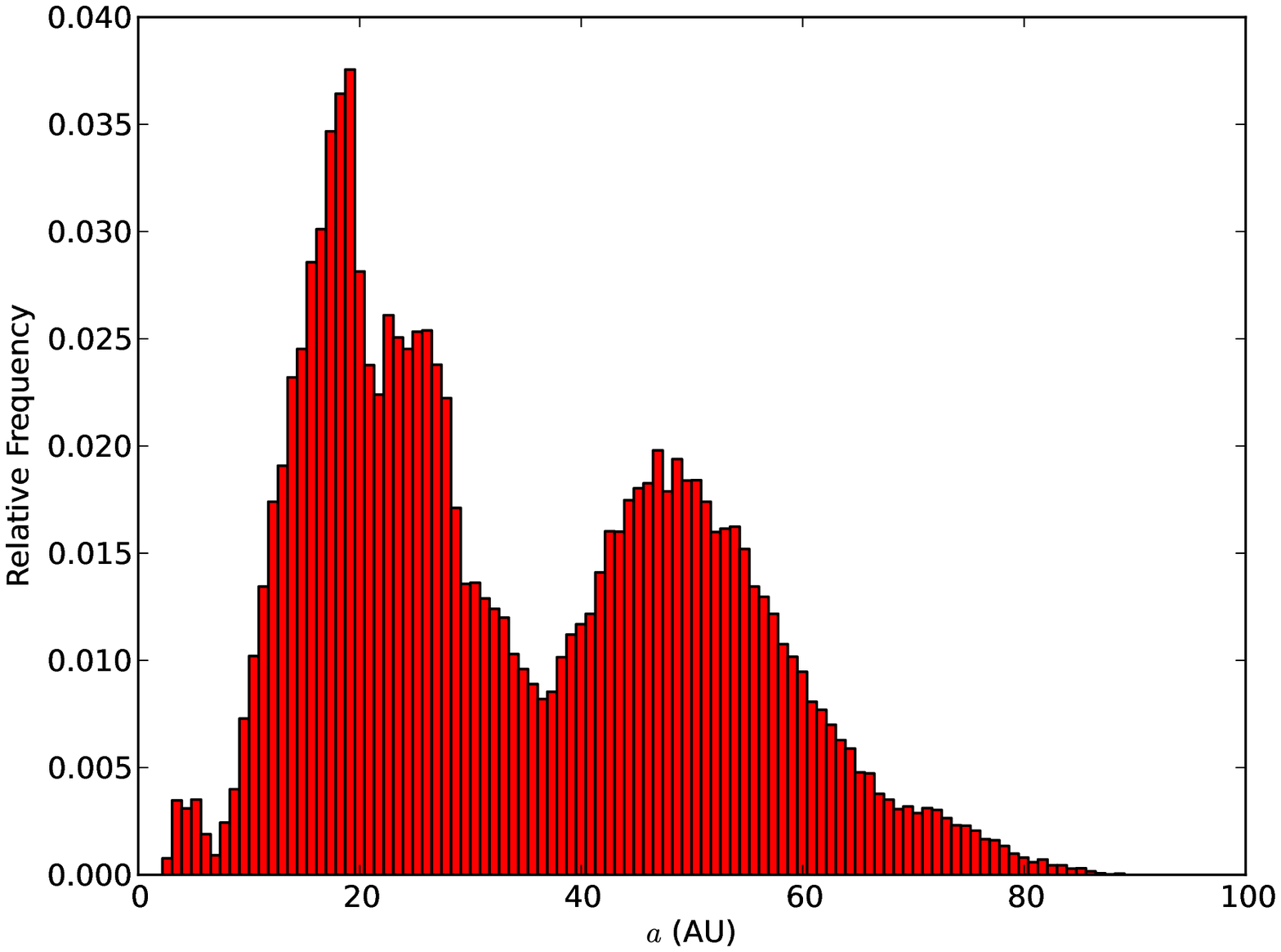} &
\includegraphics[scale=0.4]{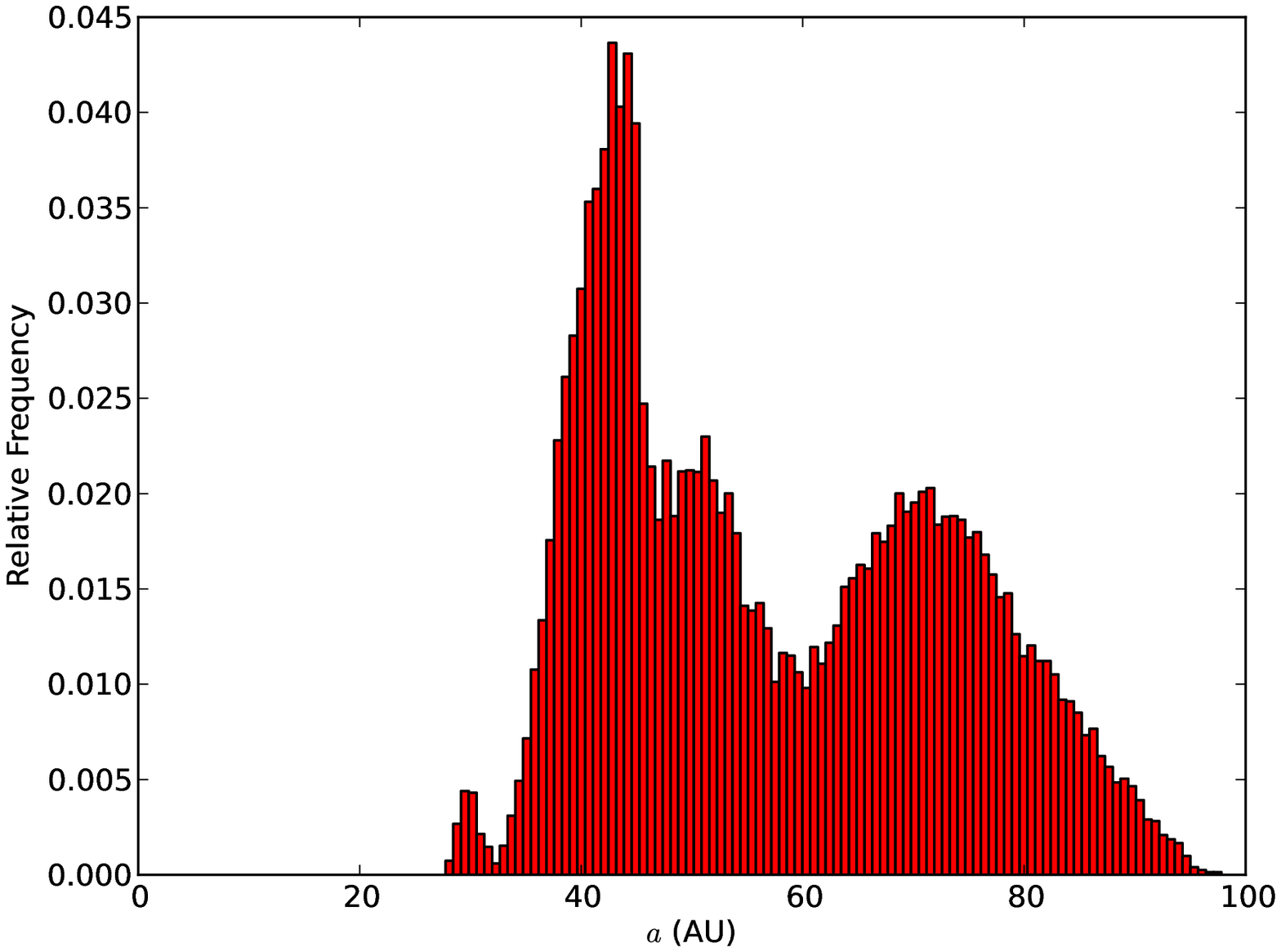} \\
\includegraphics[scale = 0.4]{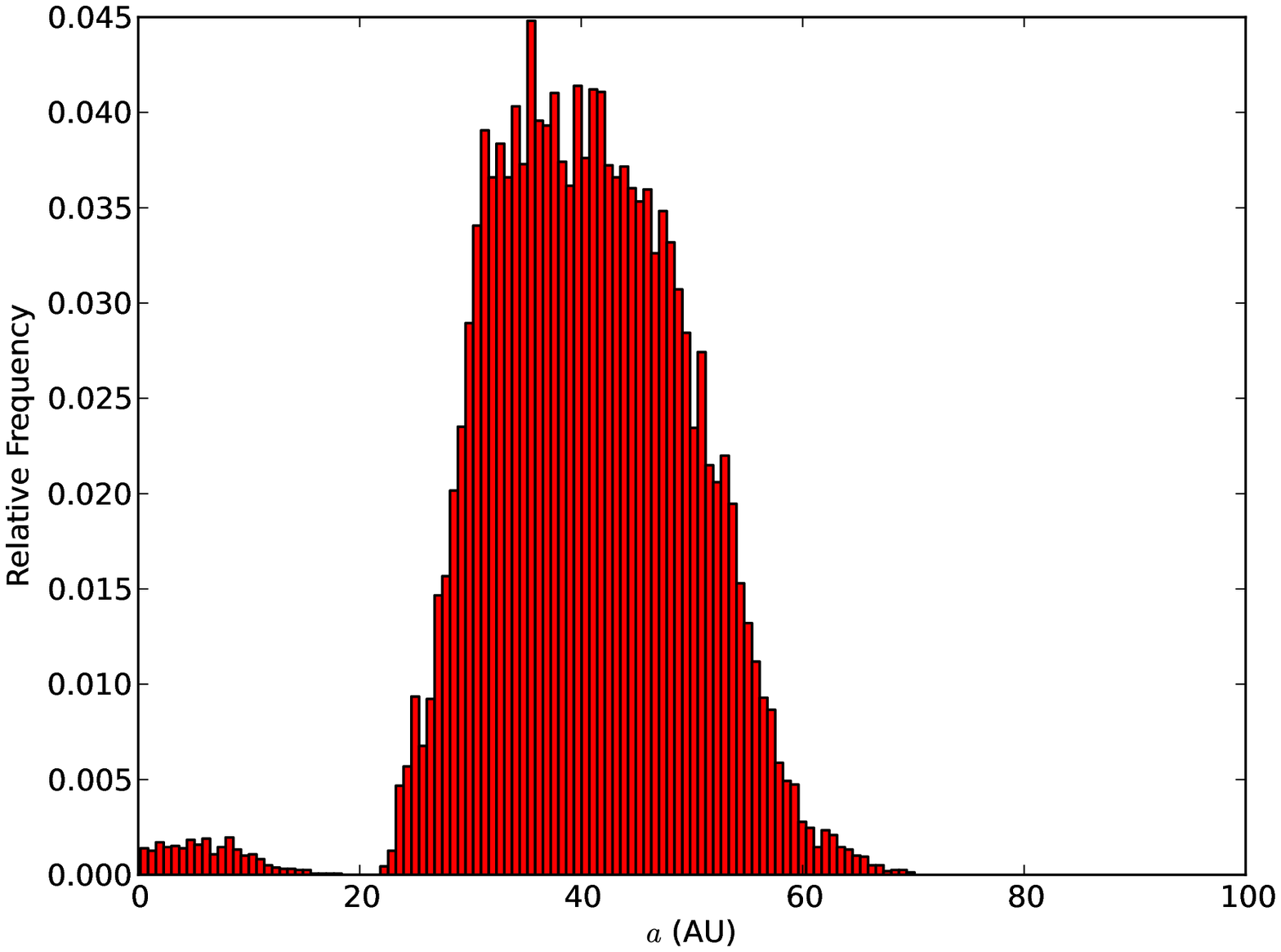} &
\includegraphics[scale=0.4]{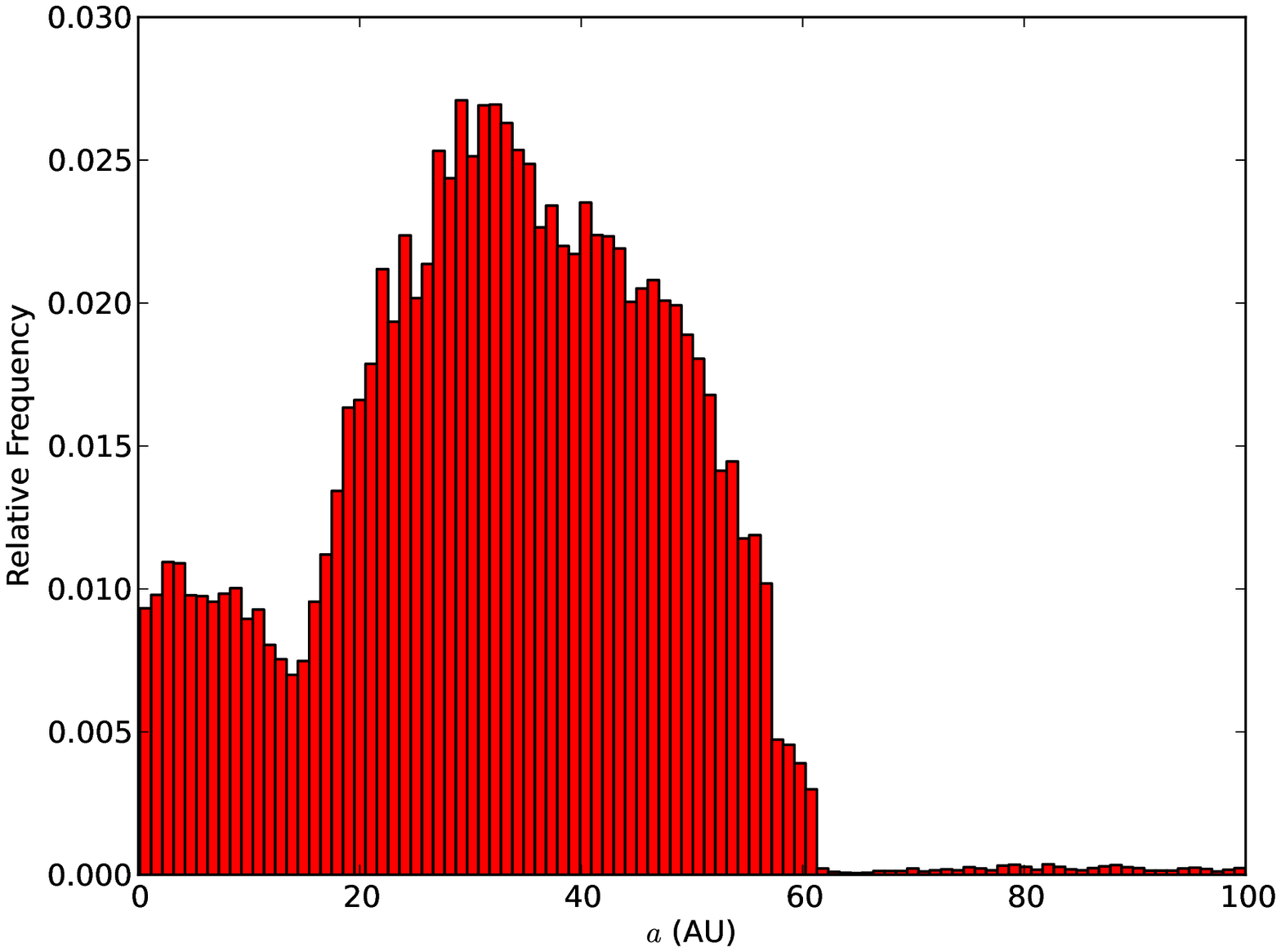} \\
\end{array}$
  \caption{The final semimajor axis distribution of all brown dwarfs for the four runs carried out in this work. \label{fig:ahist_BD}}
\end{center}
\end{figure*}

\noindent The population synthesis model predominantly produces brown
dwarfs - this is true for all four runs carried out in this work.
Depending on the opacity law, as much as 90\% of the objects that
survive the tidal downsizing process satisfy the canonical $M>13\mjup$
criterion for being brown dwarfs.  Does the brown dwarf desert
\citep{Marcy2000,Halbwachs2000} survive the migration process? In
Figure \ref{fig:ahist_BD} we plot the semimajor axis distribution of
the brown dwarfs formed in all four runs.

If we define the desert as a lack of objects below around 5 au, then
the runs with $p_{\kappa}=1$ and disc truncation show this lack quite
strikingly.  If migration is suppressed (top right of Figure
\ref{fig:ahist_BD}), then this desert extends as far as the
fragmentation boundary of the disc model.  Changing the opacity law or
not truncating the disc makes the desert less striking - brown dwarfs
are then able to survive to the inner edge of the disc model at 0.1
au.  The $p_{\kappa}=2$ run still preserves a desert in between $0.2 <
r< 20$ au, but the run without disc truncation shows a desert with
greatly reduced significance.

This would suggest that for a strong brown dwarf desert to be formed, the
disc should truncate during the fragmentation process (or migration
should be inefficient), and the embryo's opacity should not increase
too quickly with temperature.  We do not fully model the so-called
``opacity gap'' normally seen at temperatures between $\sim$ 500 -5000
K (e.g. \citealt{Bell_and_Lin}), which would significantly affect the
collapse of the embryo envelope.  A new set of grain growth and
sedimentation equations where $p_{\rm \kappa}$ is allowed to vary with
temperature would also significantly affect the potential for core
formation.

Also, we do not model ejection of brown dwarfs via scattering or three-body
interactions \citep{Bate_cluster_03,Goodwin2007}, which may be more
efficient at removing coreless objects than objects with cores
\citep{Liu2012}.  Regardless, our results would suggest that the
destruction of inwardly migrating brown-dwarf-mass objects is already
quite efficient at removing close separation brown dwarf companions
\citep{Armitage2002}, and that the restriction of disc fragmentation
to semimajor axes well above 5 au would ensure those brown dwarfs
which survive disruption would be unlikely to form close separation
binaries with their parent star.

\section{Conclusions }\label{sec:conclusions}

\noindent We have presented the first attempt at a population
synthesis model for objects produced by self-gravitating disc
fragmentation and subsequent tidal downsizing.  The model imposes a
protoplanetary disc, which can then fragment under gravitational
instability.  The fragments can then migrate, grow grains, sediment
those grains to the centre of the fragment and form solid cores, or be
tidally disrupted (not necessarily in that order).  Each
protoplanetary disc (if it fragments) produces several objects -
running the model many times with different disc properties produces a
large number of objects, allowing the construction of statistics on
tidal downsizing as a planet formation mode.

We ran the population synthesis model for four different scenarios,
varying the speed of migration, how the opacity within the embryo changes with
temperature, and whether the fragmentation process truncates the disc
or not.  Each run simulates a sufficient number of discs to produce
$\sim 10^5$ fragments. 

In general, we find that around half of all fragments are completely
destroyed during the tidal downsizing process.  The majority of
surviving fragments remain at semimajor axes $>20$ au after 1 Myr of
evolution.  Their masses are typically greater than 5 $\mjup$, with a
large fraction exceeding the brown dwarf mass limit of 13 $\mjup$.
Solid cores form with a distribution of masses peaked typically at 6
$\mearth$.  Only a very small fraction of these solid cores lose their
outer envelopes to form terrestrial planets ($<0.001$\%).  We see no
production of planetesimal belts via evolved grain populations
becoming unbound during the destruction of the embryo (although we saw some weak evidence of processed solids being returned to the disc via disruption).

The model process does require further development, and the physics to be added will both suppress and enhance grain growth/core formation.  More appropriate migration models may speed up the tidal disruption process \citep{Baruteau2011}, but equally more faithful modelling of enrichment at birth \citep{Boley2010} planetesimal capture \citep{Helled2006} and opacity evolution during embryo contraction \citep{Helled2011} can produce cores much more rapidly \citep{Helled2008a,Helled2011}.  In short, it remains possible that the gravitational instability mode of planet formation can form a modest number of terrestrial planets, but it appears to form a much larger quantity of giant planets and brown dwarfs that preferentially exist at semimajor axes above around 30 AU.

\section*{Acknowledgments}

\noindent DF and KR gratefully acknowledge support from STFC grant
ST/J001422/1.  The authors wish to thank Sergei Nayakshin for useful
discussions regarding the construction of this model, and the referee for their comments which greatly strengthened and balanced this paper.

\bibliographystyle{mn2e} 
\bibliography{TD_synthesis}

\appendix

\label{lastpage}

\end{document}